\documentclass[reprint,amsmath,amssymb,aps]{revtex4-2}

\usepackage{graphicx}
\usepackage{dcolumn}
\usepackage{bm}

\usepackage{subcaption}
\usepackage{float}

\begin{document}

\preprint{APS/123-QED}

\title{Braneworld Neutron Stars: Constraining Brane Tension with Observational Data}
\author{Masum Murshid}
\email{masum.murshid@wbscte.ac.in}
\author{Nilofar Rahman}
\email{rahmannilofar@gmail.com}
\author{Mehedi Kalam}
 \email{kalam@associates.iucaa.in(corresponding author)}
\affiliation{Department of Physics, Aliah University, II-A/27, Action Area II, Newtown, Kolkata-700160, India.}%


\begin{abstract}
In this article, we investigate the properties of neutron stars within the braneworld model, employing six distinct piece-wise polytropic equation of states. These equation of states satisfy observational constraints put by GW170817 event and pulsar observations (PSR J0740 and PSR J0030) within general relativity framework. Our primary goal is to assess whether these equation of states, in conjunction with the braneworld framework, can accommodate more massive neutron stars, as suggested by the GW190814 observation, while remaining consistent with established observational constraints. The brane tension parameter significantly affects the mass-radius and mass-tidal deformability relations, particularly for neutron stars with masses exceeding the canonical value. We establish strong constraints on the brane tension by comparing the canonical neutron star radius and tidal deformability with the results from the braneworld model, a stringent lower bound on the brane tension, $\lambda > 2 \times 10^ {37} \, \text {dyne/cm} ^2 $. Our results demonstrate that the braneworld model allows for the existence of neutron stars with masses greater than those predicted by General Relativity, in agreement with the GW190814 observation, and highlight the significant role of brane tension in shaping the properties of neutron stars.
\end{abstract}

\maketitle

\section{Introduction}

Einstein's theory of General Relativity (GR), proposed in 1915, provides the most accurate understanding of gravity on a large scale. According to GR, spacetime is a continuum consisting of three spatial dimensions and one time dimension—matter and energy warp this spacetime, resulting in the force we recognize as gravity. Although GR has proved to be a very effective theory for explaining many astrophysical and cosmological events, it has difficulty with Quantum Field Theory (QFT), as the particle behaviour is at microscopic scales here. This conflict between GR and QFT has been a central area for many years, leading to efforts to unify them, such as string theory and loop quantum gravity. The concept of extra dimensions is one of the concepts that has gained interest in this concept. In 1919, Kaluza \cite{Kaluza:1921} wrote to Einstein suggesting that gravity and electromagnetic could be combined into a single geometric theory using a 5D theory of GR.

The critical query is: If additional dimensions exist, then why can't we detect them? One explanation might be that only gravity, in contrast to other forces, can expand into these additional dimensions. In particle physics, this is referred to as the hierarchy problem. Several approaches have been proposed to address this hierarchy problem. The Randall-Sundrum (RS) model is one of the practical approaches that can solve this problem.

RS's first braneworld model \cite{Randall:1999a} proposed a framework consisting of two branes: one with positive tension and the other with negative tension. In this model, the positive tension brane is identified as the one representing our universe. Within the higher dimensional bulk of this model, the (3+1) dimensional branes are embedded. All the forces are restricted within the brane, but gravity is the only force extending into the bulk. The second brane is eliminated in the RS2 model \cite{Randall:1999b}, which maintains the wrapedness but only considers a single Planck brane with an extended additional dimension. The metric is the same as in RS1. While the gravitational force is not infinitely weak due to the endless nature of the extra dimension, the wrapping process ensures that gravity remains concentrated near the brane. In contrast to RS1, the RS2 model focuses on how gravity may function in higher dimensions and still produce Newtonian gravity at observable sizes rather than directly addressing the hierarchy problem. Since the energy hierarchy can be explained without introducing a second brane, the RS2 model is very appealing for investigating the nature of extra dimensions and gravity.

In the RS2 model, Shiromizu et al. \cite{Shiromizu:2000} derived field equations on the brane using the Gauss-Codazzi equations, establishing matching conditions and applying \(Z_{2}\) symmetry. Their foundational work demonstrated that standard Einstein's field equations are modified by additional terms that account for bulk effects on the brane. The bulk corrections to the Einstein equations can be categorized into two types. First, local quadratic energy-momentum corrections are introduced by matter fields. Second, non-local effects arise from the free gravitational field in the bulk, which are communicated through the projection of the electric part of the bulk Weyl tensor. The influence of local energy-momentum corrections becomes significant only at extremely high energy levels. The non-local effects can be further decomposed into non-local energy density, flux, and anisotropic stress components \cite{Maartens:2000}. The asymptotic behaviour of the electric part of the Weyl tensor for the localized matter perturbation indicates that it decreases sufficiently fast not to affect conserved quantities at spatial infinity, resulting in the holding of usual conservation laws, such as ADM energy conservation, on the brane in asymptotically flat spacetime \cite{Sasaki:2000}. The exact solution for static blackholes localized on 3-brane within the context of the 5D RS2 braneworld scenario is the Reissner-Nordstrom metric, which represents a blackhole without electric charge but with the tidal charge from 5D. This tidal charge correction is a change made to the Schwarzschild potential based on new field equations of braneworld. This correction is negative and is not allowed in general relativity. It creates a horizon located at a greater radius than the Schwarzschild horizon, describing the robust gravity regime on the brane \cite{Dadhich:2000}. No unique solution is found for non-vanishing exterior Weyl energy and pressure even after the assumption of spacetime is asymptotically Schwarzchild \cite{Germani:2001}. Several interior and exterior exact solutions of the compact stars are found in \cite{Ovalle:2008, Ovalle:2009, Ovalle:2010, Ovalle:2013, Casadio:2012}. The spherically symmetric star or blackhole solutions were investigated from the perspective of the bulk in \cite{Creek:2006}. The cosmological aspects of braneworld scenarios were investigated in several studies, including those by \cite{Binetruy:1999, Maeda:2000, Maartens:2000, Langlois:2001, Chen:2001, Kiritsis:2005, Campos:2001}.

Castro et al. \cite{Castro:2014} studied the macroscopic properties of compact objects by solving the Tolman-Oppenheimer-Volkoff equations. They examined three potential compact objects: hadronic, hybrid, and quark stars. In their analysis, they assumed a simple relationship between the Weyl energy density ($\mathcal{U}$) and pressure ($\mathcal{P}$), expressed as $\mathcal{P} = \omega \mathcal{U}$, putting a lower limit restriction on the brane tension i. e. $\lambda \ge 3.89 \times 10^{36} \text{dyne}/ \text{cm}^{2}$ by demanding at least a $1.44 M_{\odot}$ star can be obtained. In a uniformly dense stellar structure where both Weyl pressure and energy density are zero, it has been demonstrated that brane corrections become significant if $\lambda \lesssim 10^{4} \, \text{MeV}/\text{fm}^{3} \, = 1.6 \times 10^{36} \text{dyne}/ \text{cm}^{2}$. Under these conditions, the star is less compact compared to what would be predicted by GR \cite{Felipe:2016}. While considering a non-vanishing Weyl energy density with $\omega = 0$ in the star's interior, the Schwarzschild solution that describes the star's exterior leads to a more compact stellar object than GR. This increased compactness occurred because the Israel-Darmois matching condition requires Weyl's energy density to be negative inside the star's interior. This negative energy density leads to uniformly dense polytropic fluid stellar objects being more compact than in GR \cite{Linares:2015}. Following the paper by Linares et al. \cite{Linares:2015}, Lugones and Arbanil demonstrated that it is possible to obtain compact objects with arbitrarily large masses \cite{Lugones:2017}.

Tidal deformability is crucial for understanding dense matter properties in binary neutron star (BNS) systems. It measures how much a neutron star deforms during a merger due to the gravitational influence of its companion star. The amount of this deformation is directly connected with the ultra-dense matter Equation of State (EoS), which controls the internal structure of the neutron star. The detection of a BNS merger, GW170817 \cite{Abbott:2017, Abbott:2018}, by the LIGO and Virgo collaborations, highlighted the importance of tidal deformability. This event made the measurement of the tidal deformability parameter ($\Lambda$) possible. The discovery of the compact object merger GW190814 \cite{Abbott:2020} has raised new challenges regarding the composition of ultra-dense matter because the secondary object in this system was either a light Blackhole or a very massive NS. Tidal deformability is associated with the Love Number ($k_2$), which depends on the mass and radius of NS.

In this article, we aim to constrain the brane tension parameter by utilizing the GW170817 event along with pulsar observations (specifically PSR J0740 and PSR J0030). This paper is organized as follows: After the introduction in Section I, we present the mathematical framework for deriving the TOV equations from the field equations within the context of the braneworld scenario in Section II. Section III is dedicated to the discussion of tidal deformability, while Section IV outlines the equations of state employed in this study. We analyze our findings in Section V and provide concluding remarks in Section VI.

\section{Field Equations on the 3-Brane world}\label{S2}
In the braneworld scenario, all the matter forces except gravity are confined within our 4D observable universe known as brane, and this brane is assumed to be embedded in a 5D spacetime known as bulk. The influence of five-dimensional gravity results in modifying Einstein's field equations within our 4D observable universe or brane. This modification can be succinctly articulated as follows:
\begin{equation}\label{S2EQ:1}
G_{\mu\nu}=8\pi T_{\mu\nu}^{\textrm{eff}}-\Lambda g_{\mu\nu}
\end{equation}
where $G_{\mu\nu}$ is the conventional Einstein's tensor and $\Lambda$ is the cosmological constant on the brane and
\begin{equation}\label{S2EQ:2}
T_{\mu\nu}^{\textrm{eff}}=T_{\mu\nu}+\frac{6}{\lambda} S_{\mu\nu}-\frac{1}{\kappa^2}\mathcal{E}_{\mu\nu}
\end{equation}
where $\lambda$ is the brane tension, corresponding to the brane's vacuum energy density. The term $S_{\mu\nu}$ in Eq.\ref{S2EQ:2} represents local bulk correction, and the term $\mathcal{E}_{\mu\nu}$ is the projection of bulk Weyl tensor represents non-local bulk correction in $4D$ world. In the natural unit $G=c=1$, the value of $\kappa^{2}$ is $8\pi$. 

In the case of a perfect fluid or minimally coupled scalar field, we can write \cite{Maartens:2000,Maartens:2010}
\begin{equation}\label{S2EQ:3}
T_{\mu\nu}=\rho u_{\mu}u_{nu}+ P h_{\mu\nu}
\end{equation}

\begin{equation}\label{S2EQ:4}
 S_{\mu\nu} = \frac{1}{12} \rho^{2} u_{\mu} u_{\nu}  + \frac{1}{12}\rho (\rho+2 P)h_{\mu\nu}
\end{equation}

Where $\rho=\rho(r)$, $P=P(r)$ and $u^{\mu}$ are, respectively, the pressure, density and four-velocity of the stellar matter of interest and $h_{\mu\nu}=g_{\mu\nu}+u_{\mu}u_{\nu}$ gives the projection orthogonal to the four-velocity. Now, if we assume a static spherically symmetric scenario, then we have
\begin{equation}\label{S2EQ:5}
\mathcal{E}_{\mu\nu}=-\frac{6}{\kappa^{2} \lambda} \left[ \mathcal{U}u_{\mu}u_{\nu}+\mathcal{P}u_{\mu}u_{\nu} +\frac{(\mathcal{U}-\mathcal{P})}{3}h_{\mu\nu}\right]
\end{equation}
where $r_{\mu}$ is the unit radial vector, $\mathcal{U}$ and $\mathcal{P}$ are, respectively, the non-local energy density, commonly known as "dark radiation" and non-local pressure, commonly known as "dark pressure"  on the brane \cite{Ovalle:2013}.

We consider stationary spherically symmetric spacetime written in the form of line elements as
\begin{align} \label{S2EQ:6}
ds^{2}&= -e^{\nu}dt^{2} +\frac{1}{1-\frac{2m}{r}}dr^2 +r^2 [d\theta^2+\sin^{2}\theta d\phi^2]
\end{align}

The modified TOV equations for the structure of the star on the brane are then expressed as:

\begin{equation}\label{S2EQ:7}
\frac{dm}{dr}=\frac{1}{2} \kappa ^2 r^2 \rho _{\text{eff}}
\end{equation}

\begin{equation}\label{S2EQ:8}
\frac{dP}{dr}=-(\rho+P)\frac{2m+ \kappa^2 \lambda  r^3 \left( P_{\text{eff}}+\frac{4 \mathcal{P}}{\kappa^2 \lambda}\right)}{ 2 r (r-2 m)}
\end{equation}

\begin{equation}\label{S2EQ:9}
\frac{d\nu}{dr}=\frac{2m+ \kappa^2 \lambda  r^3 \left( P_{\text{eff}}+\frac{4 \mathcal{P}}{\kappa^2 \lambda}\right)}{ r (r-2 m)}
\end{equation}

\begin{align}\label{S2EQ:10}
\frac{d\mathcal{U}}{dr}=& -2 \frac{d\mathcal{P}}{dr}-\frac{6
   \mathcal{P}}{r}\\ \nonumber
  & +\frac{2m+ \kappa^2 \lambda  r^3 \left( P_{\text{eff}}+\frac{4 \mathcal{P}}{\kappa^2 \lambda}\right)}{ r (r-2 m)}\left[\frac{\kappa ^4  (P+\rho )^2}{4 v_{s}^{2} }- (\mathcal{P}+2 \mathcal{U})\right]
\end{align}
where $v_{s}^{2}=\frac{dP}{d\rho}$ and 
\begin{equation}\label{S2EQ:11}
\rho _{\text{eff}}= \rho+\frac{\rho^2}{2\lambda}+\frac{6}{\kappa^4 \lambda}\mathcal{U}
\end{equation}

\begin{equation}\label{S2EQ:12}
P_{\text{eff}}= P+\frac{P\rho}{\lambda}+\frac{\rho^2}{2\lambda}+\frac{2}{\kappa^4 \lambda}\mathcal{U}
\end{equation}

We need three boundary conditions to solve Eqs.\ref{S2EQ:7}-\ref{S2EQ:10}. Two of the three boundary conditions are the same as for the standard general relativistic equations, which are as follows
\begin{equation}\label{S2EQ:13}
m(r=0)=0 \hspace{0.7cm}  \text{and} \hspace{0.7cm} P(R)=0
\end{equation}

To determine other remaining boundary conditions, we use Israel-Darmois matching condition $\left[G_{\mu\nu}r^{\nu}\right]_{\Sigma}=0$ at the star's surface $\Sigma$, where $\left[f\right]_{\Sigma}=f(R^{+})-f(R_{+})$. Applying this  Israel-Darmois matching condition on the brane field Eq.\ref{S2EQ:1}, for any static spherical objects, we get
\begin{equation}\label{S2EQ:14}
\frac{\kappa^{2} \rho^{2}(R)}{4}+\mathcal{U}^{-}(R)+2\mathcal{P}^{-}(R)=\mathcal{U}^{+}(R)+2\mathcal{P}^{+}(R)
\end{equation}

Following the paper \cite{Lugones:2017}, we first assume the exterior spacetime to be the Schwarzschild solution, where $\mathcal{U}^{+}=0$ and $\mathcal{P}^{+}=0$. Secondly, we assume $\mathcal{P}^{-}=0$ to maintain the isotropy of the effective stress-energy tensor. These two assumptions simplify Eq. \ref{S2EQ:14} as
\begin{equation}\label{S2EQ:15}
\frac{\kappa^{2} \rho^{2}(R)}{4}+\mathcal{U}^{-}(R)=0
\end{equation}

For a given equation of state ($\rho=\rho(P)$), to obtain the mass and radius of the star, it is necessary to solve the differential equations (Eqs.\ref{S2EQ:7}-\ref{S2EQ:10}) from the centre of the star ($r=0$) to its surface, and then from the surface to infinity, using the boundary conditions provided by Eq.\ref{S2EQ:13} and an initial value of $\mathcal{U}^{-}(r=0)=\mathcal{U}_{c}$. The value of $\mathcal{U}^{-}(r=0)$ is then adjusted using the shooting method to satisfy the third boundary condition given by Eq.\ref{S2EQ:15}. The mass of the star can be determined as:

\[ M= \lim_{r \to \infty} m(r) \]

\section{Tidal deformability}
In this section, we will focus on the binary neutron star system. When a neutron star is part of a binary system, it becomes deformed by the gravitational pull of its companion star. The level of deformation of the neutron star, known as tidal deformability, is dependent on the tidal field of its companion star, denoted as $\epsilon_{ij}$, is given by
  \begin{equation}\label{S3EQ:1}
\bar{\Lambda}=-\frac{Q_{ij}}{\epsilon_{ij}}  
  \end{equation}
  where quadrupole moment $Q_{ij}$ arises as a result of the gravitational tidal forces exerted by its companion. The dimensionless second Love number $k_{2}$ is related to the tidal deformability $\Lambda$ as
   
  \begin{equation}\label{S3EQ:2}
k_{2}=\frac{3}{2}~\frac{\bar{\Lambda}}{R^{5}}  
  \end{equation}
  and the dimensionless tidal deformability is defined as
  \begin{equation}\label{S3EQ:3}
  \Lambda=\frac{\bar{\Lambda}}{M^{5}}
  \end{equation}
  
 We study linear perturbation of the background to find tidal love number by considering Thorne \& Campolattaro method \cite{Thorne:1967} as following 
 \begin{equation}\label{S3EQ:4}
  g_{ab}=g^{0}_{ab}+h_{ab}
  \end{equation}
  where $g^{0}_{ab}$ is unperturbed metric and $h_{ab}$ is even-parity perturbation in Regge-Wheeler gauge \citep{Regge:1957} which is given by
  \begin{equation}\label{S3EQ:5}
  h_{ab}= \text{diag}[H_{0}e^{\nu},\frac{H_{2}}{1-\frac{2m}{r}}, r^{2}K, r^{2}K \sin^{2}\theta] Y_{2m}(\theta, \phi)
  \end{equation}
  where $H_{0}$, $H_{2}$ \& $K$ are functions of $r$, and $Y_{2m}(\theta, \phi)$ is the associate Legendre function. Under this perturbation, the perturbed stress-energy tensor can be written as
  \begin{equation}\label{S3EQ:6}
  \delta T^{\text{eff}}_{\mu\nu} =diag[-\delta \rho_{\text{eff}}, \delta P_{\text{eff}}, \delta P_{\text{eff}}, \delta P_{\text{eff}}]
  \end{equation}
  Following T. Hinderer formalism \citep{Hinderer:2008}, we get $H_{2}=H_{0}$ , $K^{'}=H_{0} \nu^{'}+H_{0}^{'}$ and a master equation for $H_{0}$  as
\begin{eqnarray}\label{S3EQ:7}
&\frac{d^{2}H_{0}}{dr^{2}}+\left[\frac{2}{r}+\frac{r}{r-2m}\left( \frac{2 m}{r^2} +\frac{\kappa^2 r}{2}(P_{\text{eff}}-\rho_{\text{eff}}) \right)\right]\frac{dH_{0}}{dr} \\ \nonumber
&+\left[\frac{r}{(r-2m)}\left\{\frac{\kappa^2 }{2}\left(5\rho_{\text{eff}}+9P_{\text{eff}} +\frac{P_{\text{eff}}+\rho_{\text{eff}}}{c_{s}^{2} }     \right) -\frac{6}{r^2}\right\} -\left(\frac{d\nu}{dr}\right)^{2} \right]H_{0}
\end{eqnarray}  
where
\begin{equation}\label{S3EQ:8}
c_{s}^{2}=\frac{dP_{\text{eff}}}{d\rho_{\text{eff}}}= \frac{v_{s}^{2} \left[8 \mathcal{U}+\kappa ^4 (P+\rho) (\lambda +\rho)\right]}{24 \hspace{0.5mm} \mathcal{U}  \hspace{0.5mm}v_{s}^{2}+\kappa ^4
   (P+\rho) (\lambda -3 P-2 \rho) }
\end{equation}
The asymptotic behaviour of Eq.\ref{S3EQ:7} at infinity yields 
\begin{align}\label{S3EQ:9}
H_{0}(r)= & \left(r-2 M    \right) r\hspace{0.5mm} H_{0_{-2}}\\ \nonumber
&+\left(\frac{1}{ r^{3}}+\frac{3M}{ r^{4}}+\frac{50 M^{2}}{7 r^{5}} \right) H_{0_{3}}
\end{align}
where the value of $H_{0_{-2}}$ and $H_{0_{3}}$ constants are to be determined by comparing solutions of Eq.\ref{S3EQ:7} at infinity with Eq.\ref{S3EQ:9}. Once these two constants are determined, the tidal love number can be calculated using the formula.
\begin{equation}
\bar{\Lambda} = \frac{H_{0_{3}}}{3 H_{0_{-2}}}
\end{equation}

\section{ Equation of State}
Neutron stars rank among the densest objects in the universe, with densities exceeding the nuclear saturation density, \( \rho_{0}=2.7 \times 10^{14} \, g/cm^{3} \). Despite this extraordinary density, the thermal temperature of neutron stars remains relatively low compared to their constituent particles' Fermi energy. This combination of low temperature and ultra-high density fosters conditions that enable the existence of exotic states of matter within the core of neutron stars. In addition to traditional nucleonic matter, the core may contain hyperons, pion or kaon condensates, and possibly strange quark matter, depending on the equation of state (EoS) used to model these extreme environments. Various theoretical approaches, including relativistic mean field (RMF) theory, quantum chromodynamics (QCD), and lattice QCD, are employed to derive these EoSs.

In this study, we adopt the piecewise polytropic approach for modelling neutron star EoSs, as proposed by Read et al. \cite{Read:2009}. This method offers a phenomenologically motivated framework to represent the EoS by parameterizing nuclear matter properties. The piecewise polytropic EoS is particularly advantageous because it enables the seamless matching of nuclear matter constraints at lower densities to astrophysical observations and predictions at higher densities.
We use six candidate EoSs for the highly dense core to ensure a comprehensive representation:  AP3, WFF2, ENG, MPA1, ALF2 and ALF4. These EoSs have been selected to cover a wide range of theoretical models and predictions:

\begin{itemize}

\item AP3 \cite{Akmal:1998} and WFF2 \cite{Wiringa:1988}  are derived from variational methods incorporating modern two- and three-body nuclear potentials.
\item ENG \cite{Engvik:1995} and MPA1 \cite{Muther:1987} use the relativistic Brueckner-Hartree-Fock (RBHF) theory, emphasizing relativistic corrections in nuclear interactions.
\item ALF2 and ALF4 \cite{Alford:2004} represent hybrid models incorporating quark matter based on the MIT bag model, providing insights into the possible deconfinement of quarks at extreme densities.

\end{itemize}

For the low-density crustal region of neutron stars, we utilize the SLy equation of state (EoS) \cite{Douchin:2001}, which is particularly well-suited for describing the outer layers of the neutron star. The SLy EoS is seamlessly integrated with the core EoSs through the parameterization provided by Read et al. \cite{Read:2009}, ensuring a smooth transition between the crust and core regions. This integration yields a comprehensive piecewise polytropic representation of the neutron star EoS. By employing these six core EoSs in conjunction with the SLy crust, we aim to systematically examine the macroscopic properties of neutron stars, such as their masses, radii, and tidal deformability within the framework of braneworld.

\begin{figure*}[!ht]
\centering
\begin{subfigure}{0.3\textwidth}
    \includegraphics[width=\textwidth]{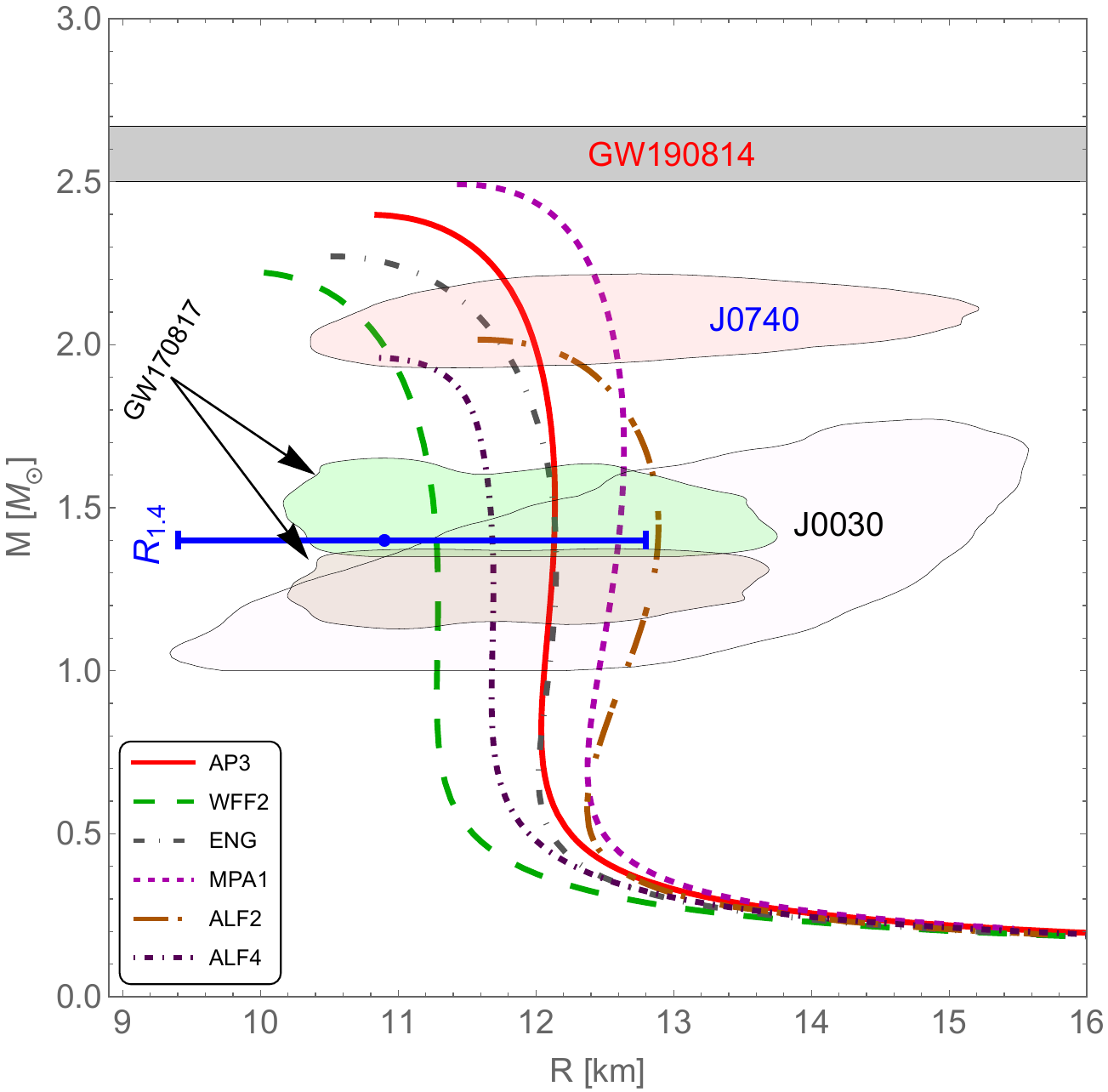}
    \caption{Mass-Radius relation}
    \label{fig:1a}
\end{subfigure}
\begin{subfigure}{0.3\textwidth}
    \includegraphics[width=0.98\textwidth]{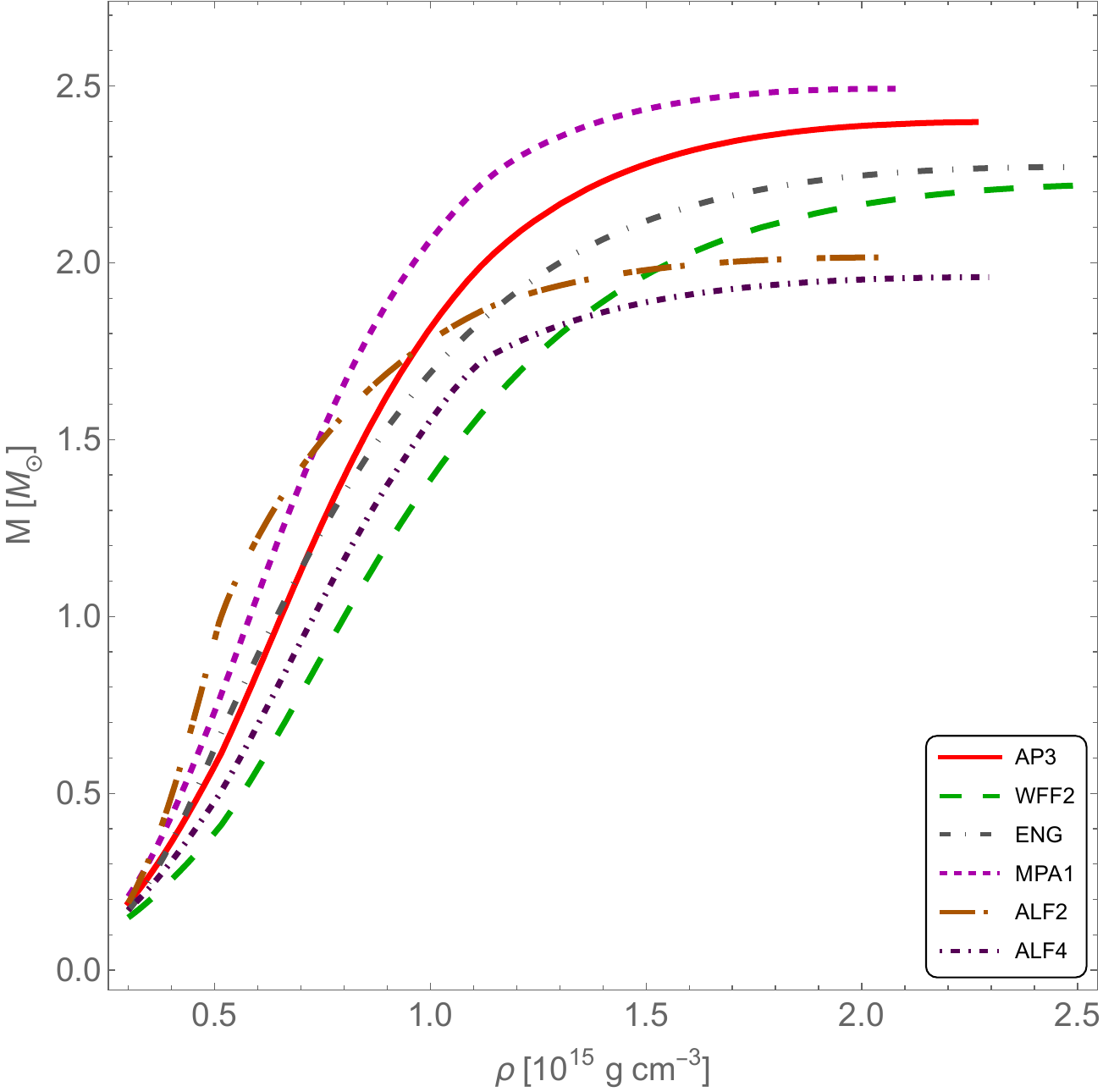}
    \caption{Mass-Density relation}
    \label{fig:1b}
\end{subfigure}
\begin{subfigure}{0.3\textwidth}
    \includegraphics[width=\textwidth]{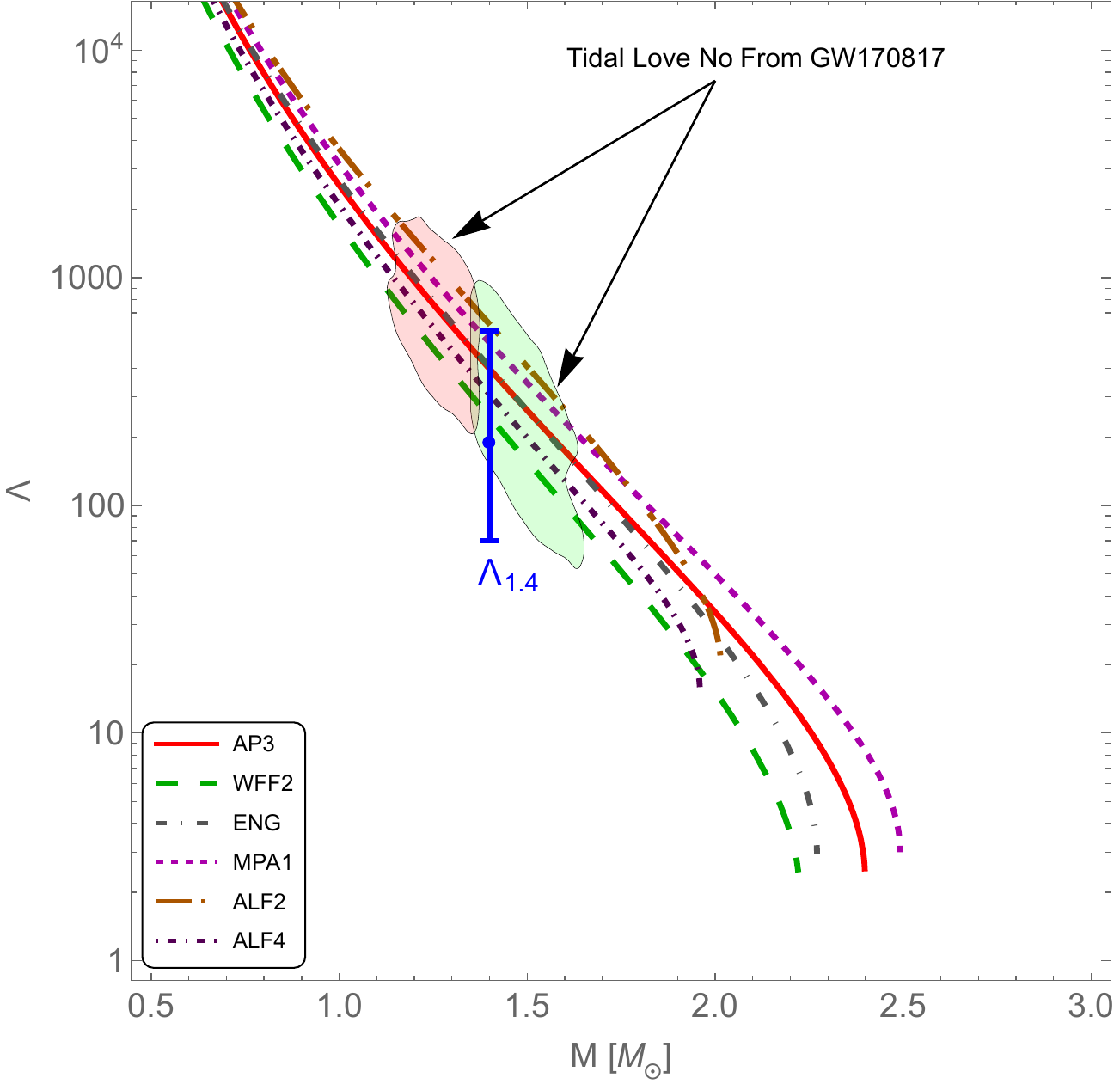}
    \caption{Mass-Love No. relation}
    \label{fig:1c}
\end{subfigure}
\caption{Relationships between the mass of compact objects and various physical properties within the framework of general relativity.}
\label{fig:1}
\end{figure*}

\begin{figure*}[!ht]
\centering
\begin{subfigure}{0.3\textwidth}
    \includegraphics[width=\textwidth]{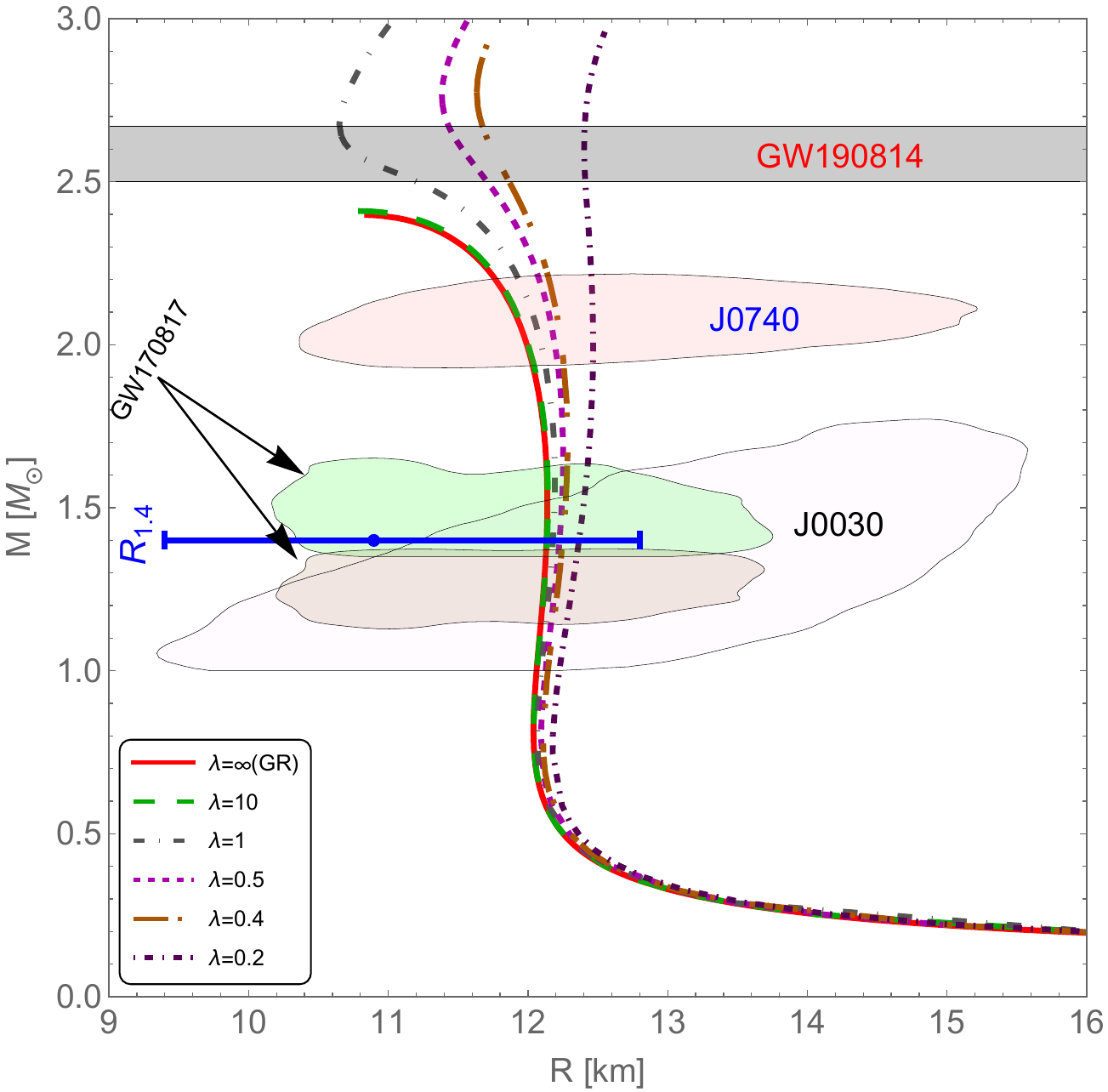}
    \caption{AP3}
    \label{fig:2a}
\end{subfigure}
\begin{subfigure}{0.3\textwidth}
    \includegraphics[width=\textwidth]{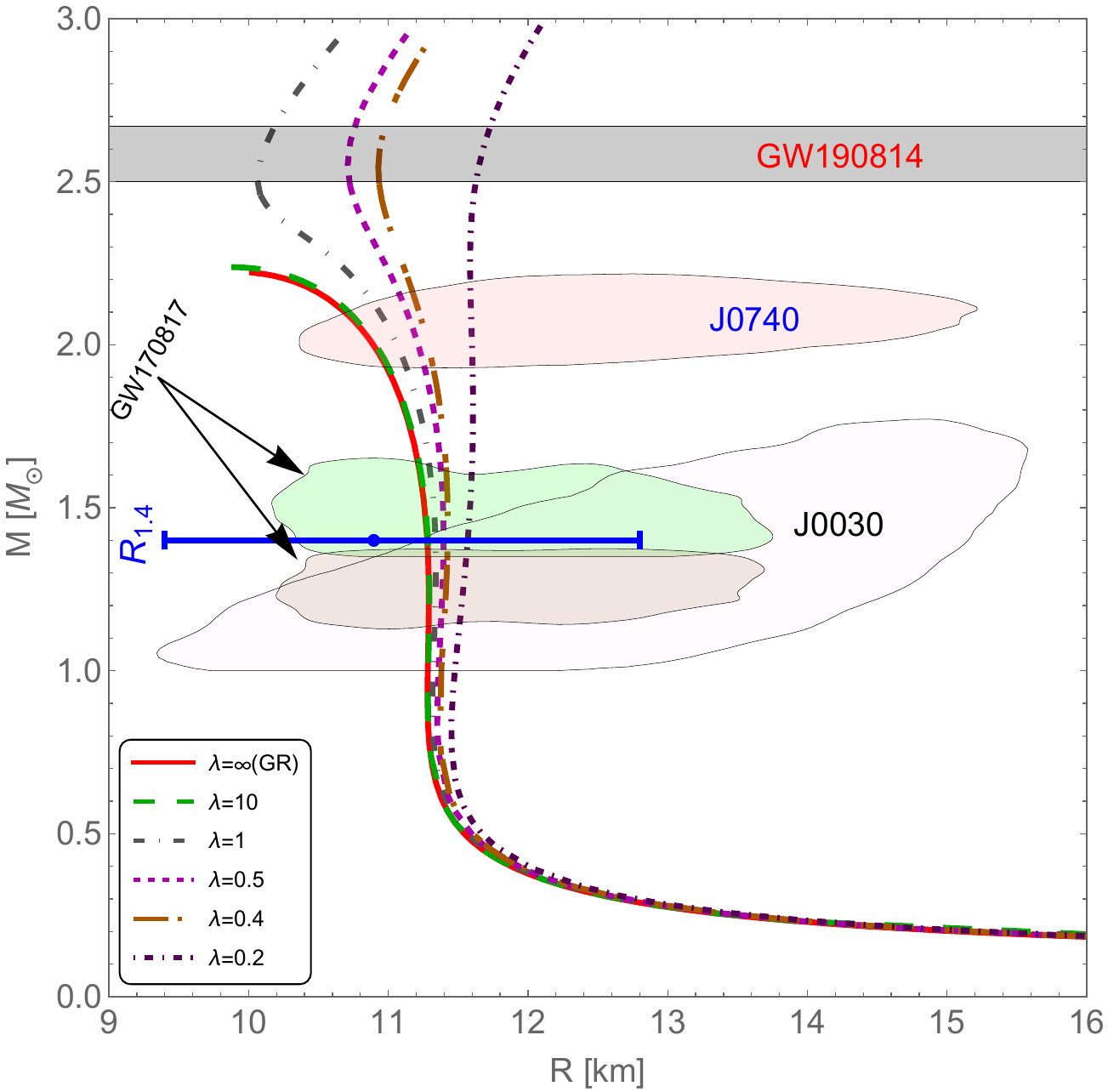}
    \caption{WFF2}
    \label{fig:2b}
\end{subfigure}
\begin{subfigure}{0.3\textwidth}
    \includegraphics[width=\textwidth]{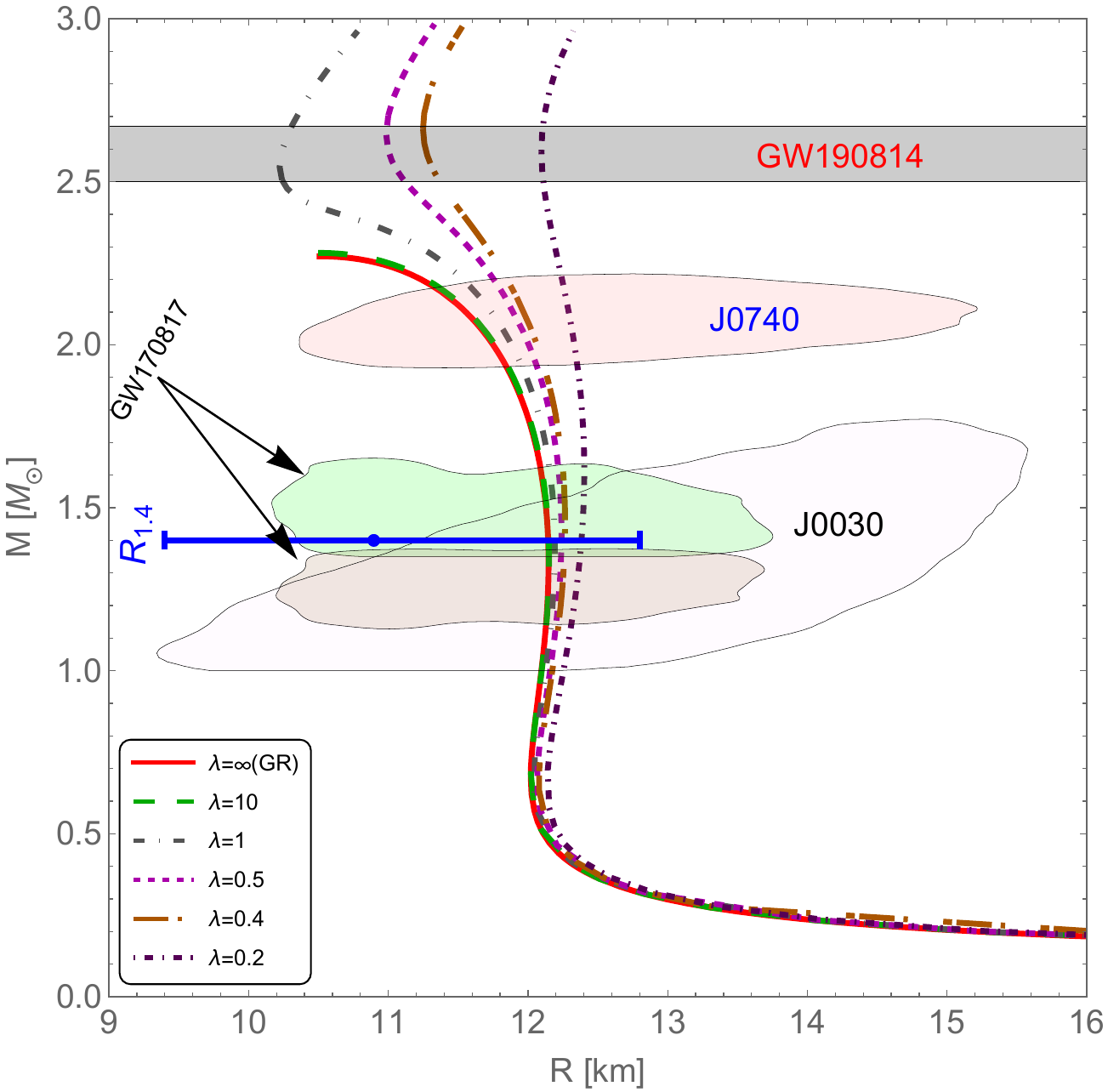}
    \caption{ENG}
    \label{fig:2c}
\end{subfigure}
\hfill
\begin{subfigure}{0.3\textwidth}
    \includegraphics[width=\textwidth]{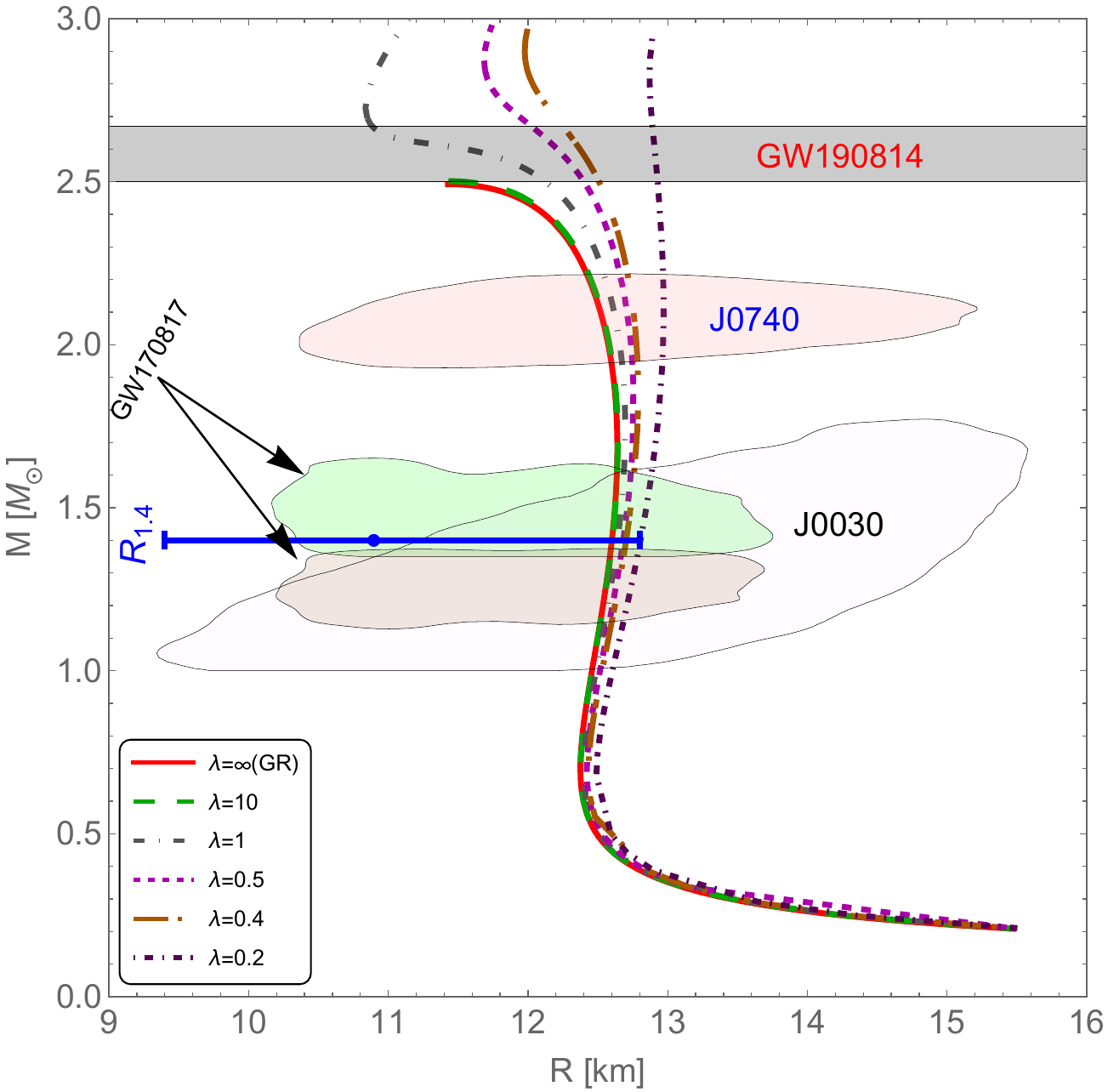}
    \caption{MPA1}
    \label{fig:2d}
\end{subfigure}
\begin{subfigure}{0.3\textwidth}
    \includegraphics[width=\textwidth]{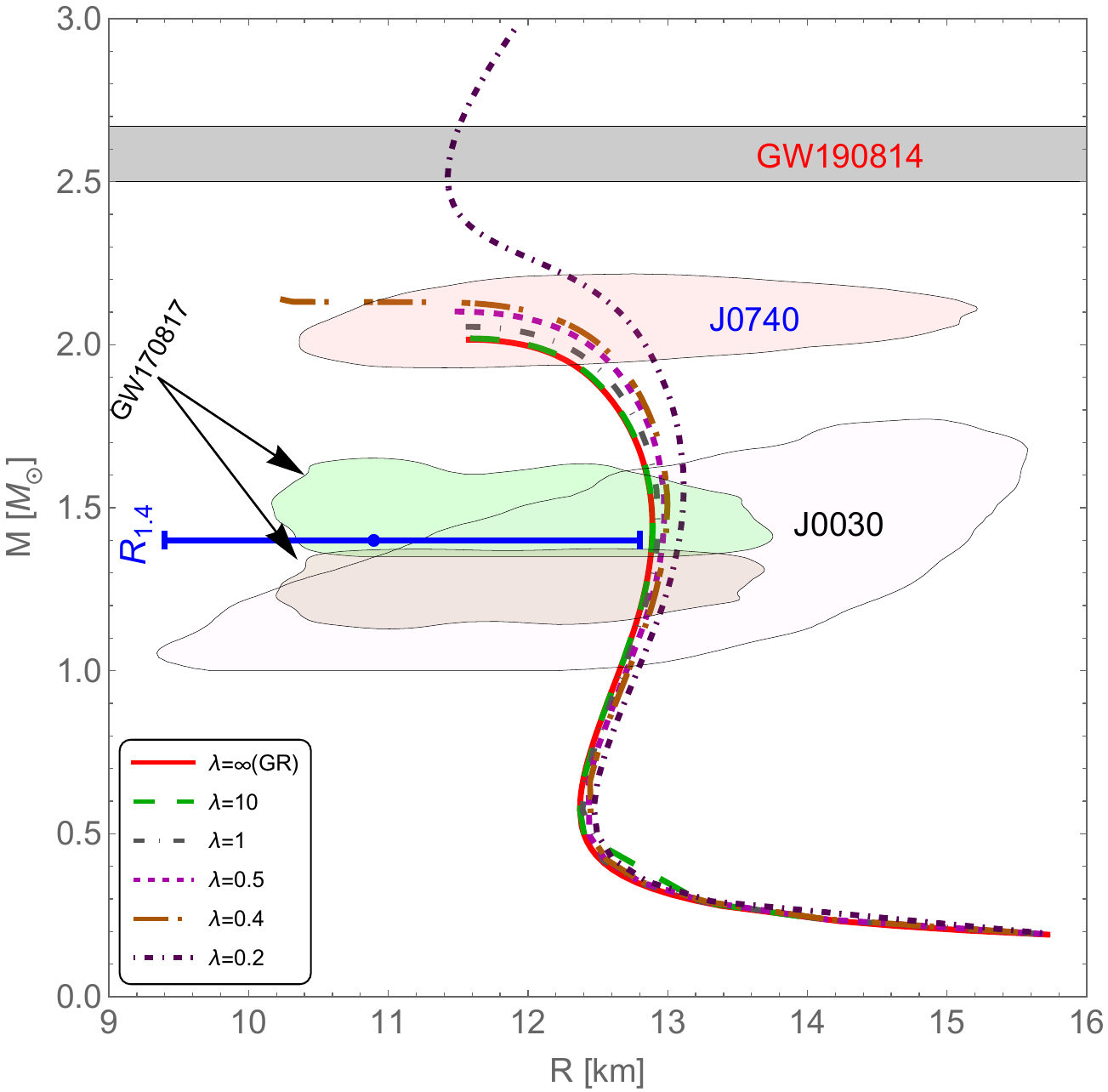}
    \caption{ALF2}
    \label{fig:2e}
\end{subfigure}
\begin{subfigure}{0.3\textwidth}
    \includegraphics[width=\textwidth]{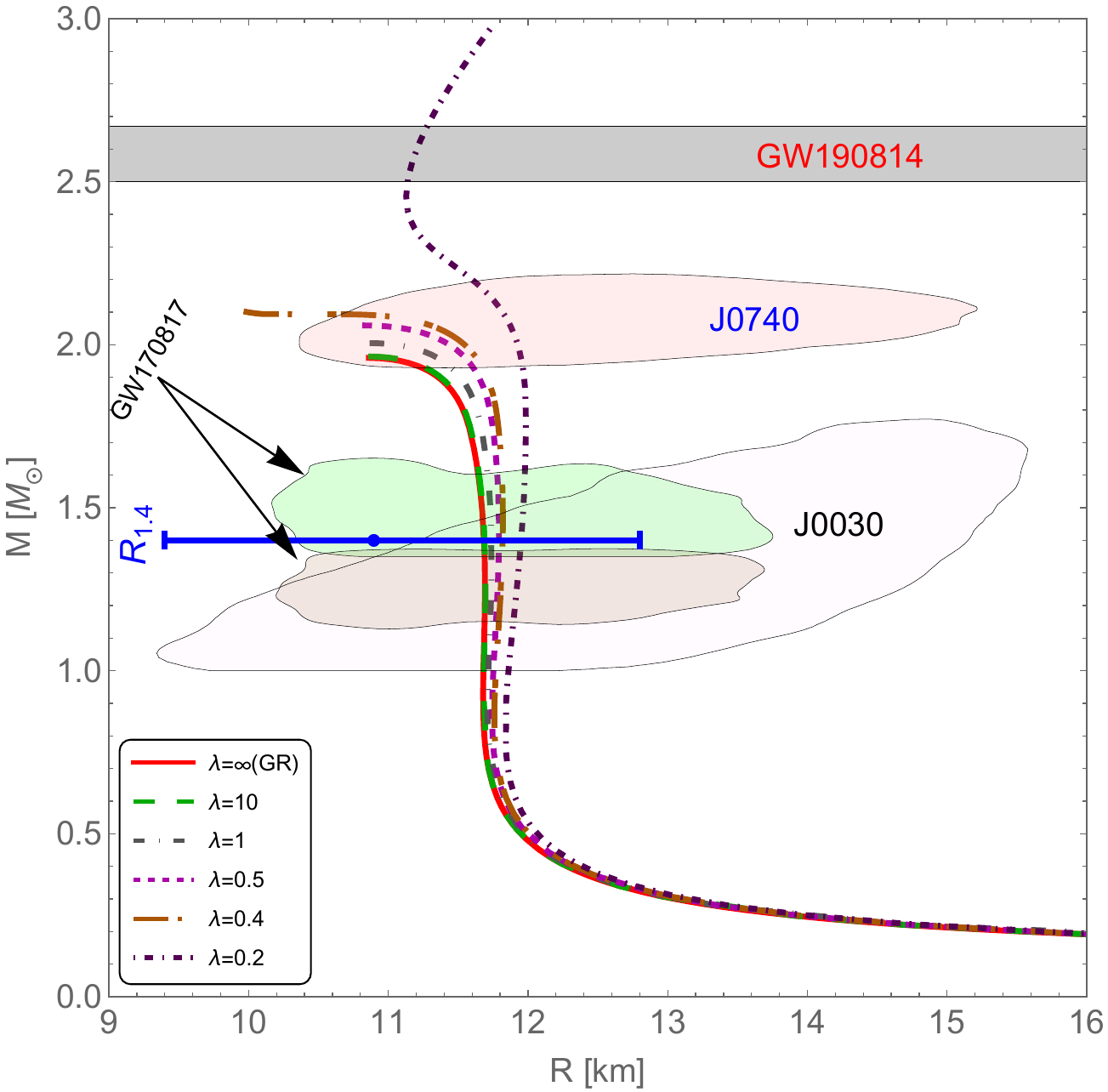}
    \caption{ALF4}
    \label{fig:2f}
\end{subfigure}
\caption{Mass-Radius relation in braneworld scenario ($\lambda$ is given in units of  $10^{38}$ $\text{dyne}/\text{cm}^2)$.}
\label{fig:2}
\end{figure*}

\begin{figure*}[!ht]
\centering
\begin{subfigure}{0.3\textwidth}
    \includegraphics[width=\textwidth]{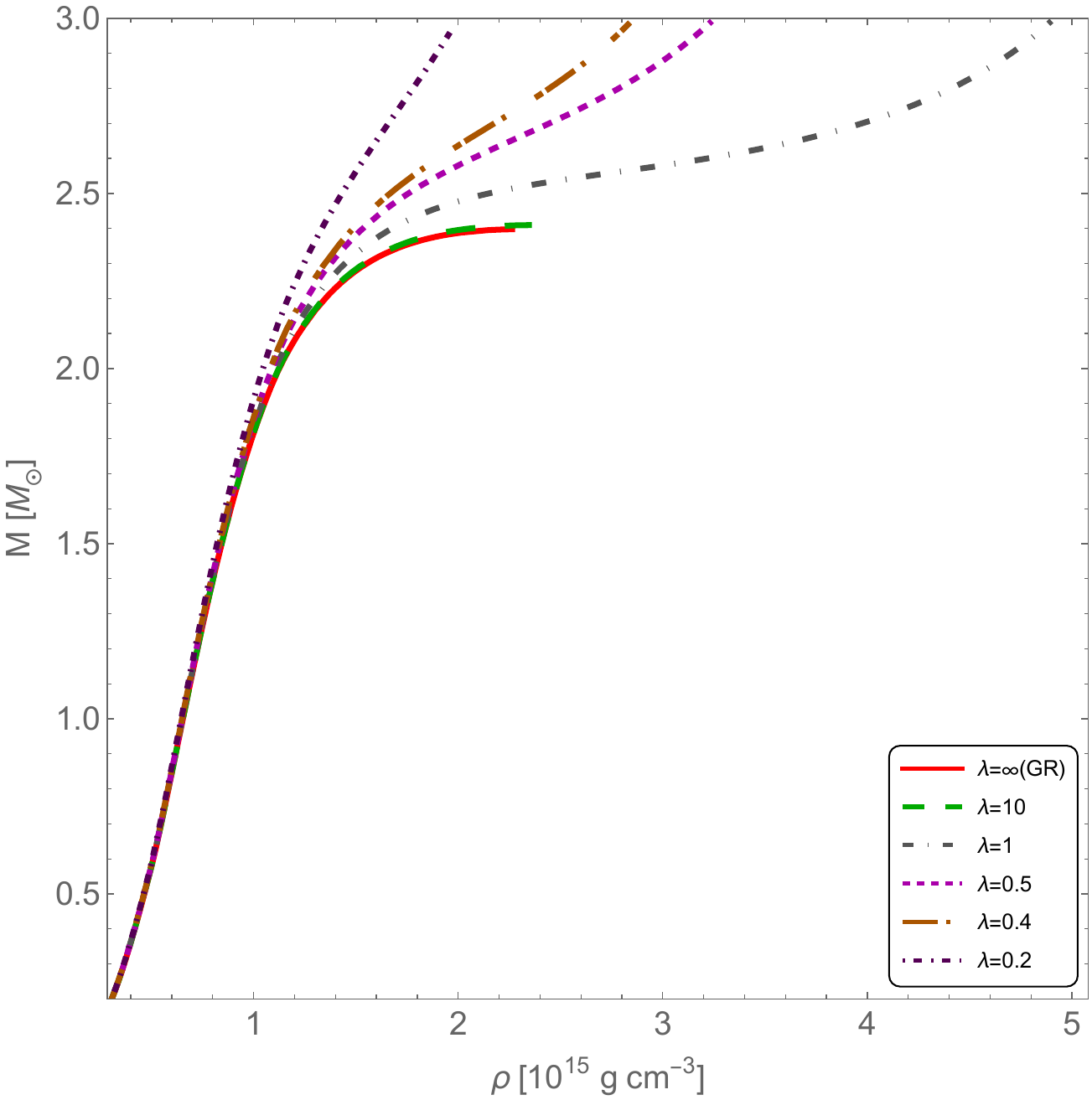}
    \caption{AP3}
    \label{fig:3a}
\end{subfigure}
\begin{subfigure}{0.3\textwidth}
    \includegraphics[width=\textwidth]{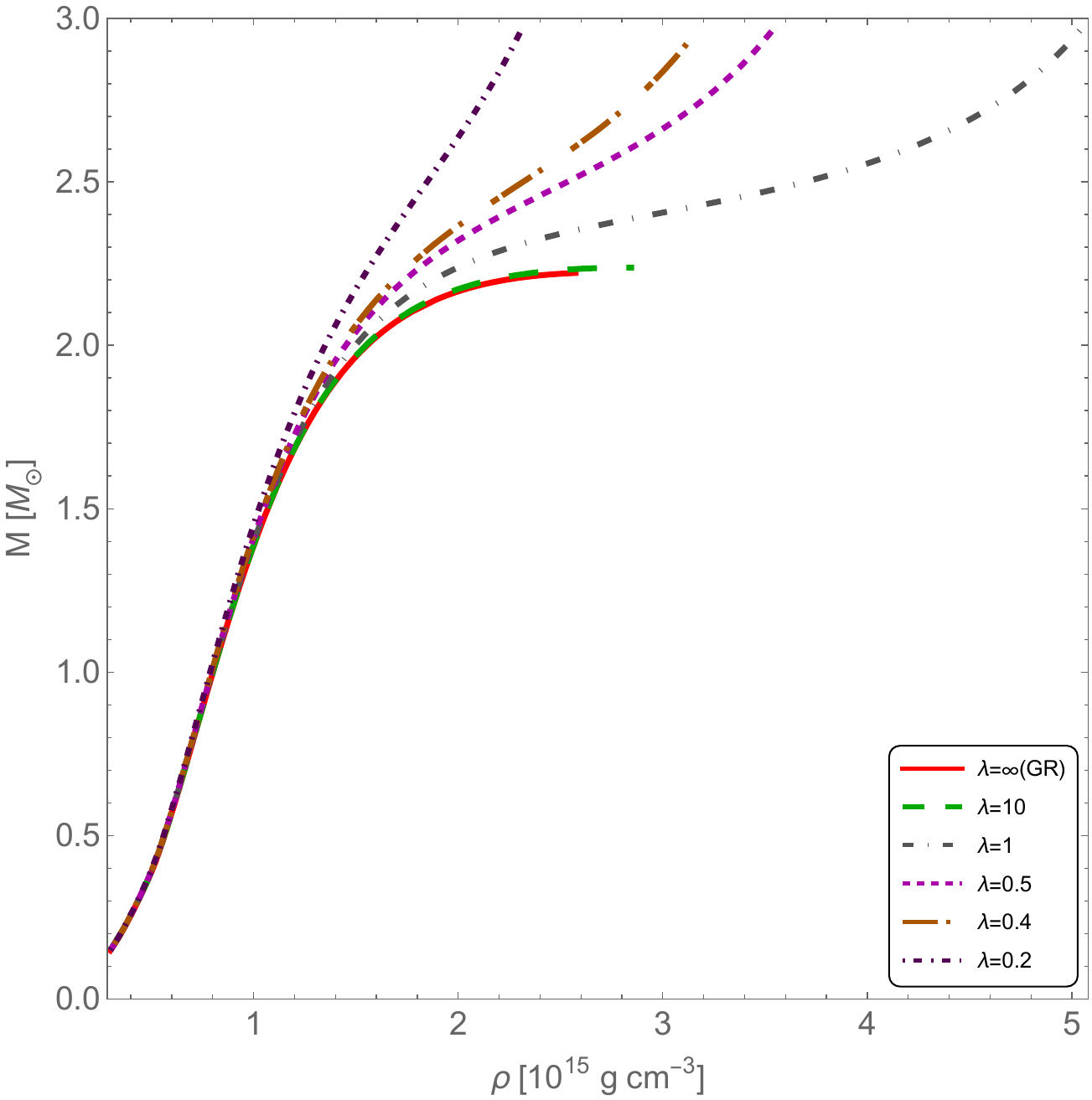}
    \caption{WFF2}
    \label{fig:3b}
\end{subfigure}
\begin{subfigure}{0.3\textwidth}
    \includegraphics[width=\textwidth]{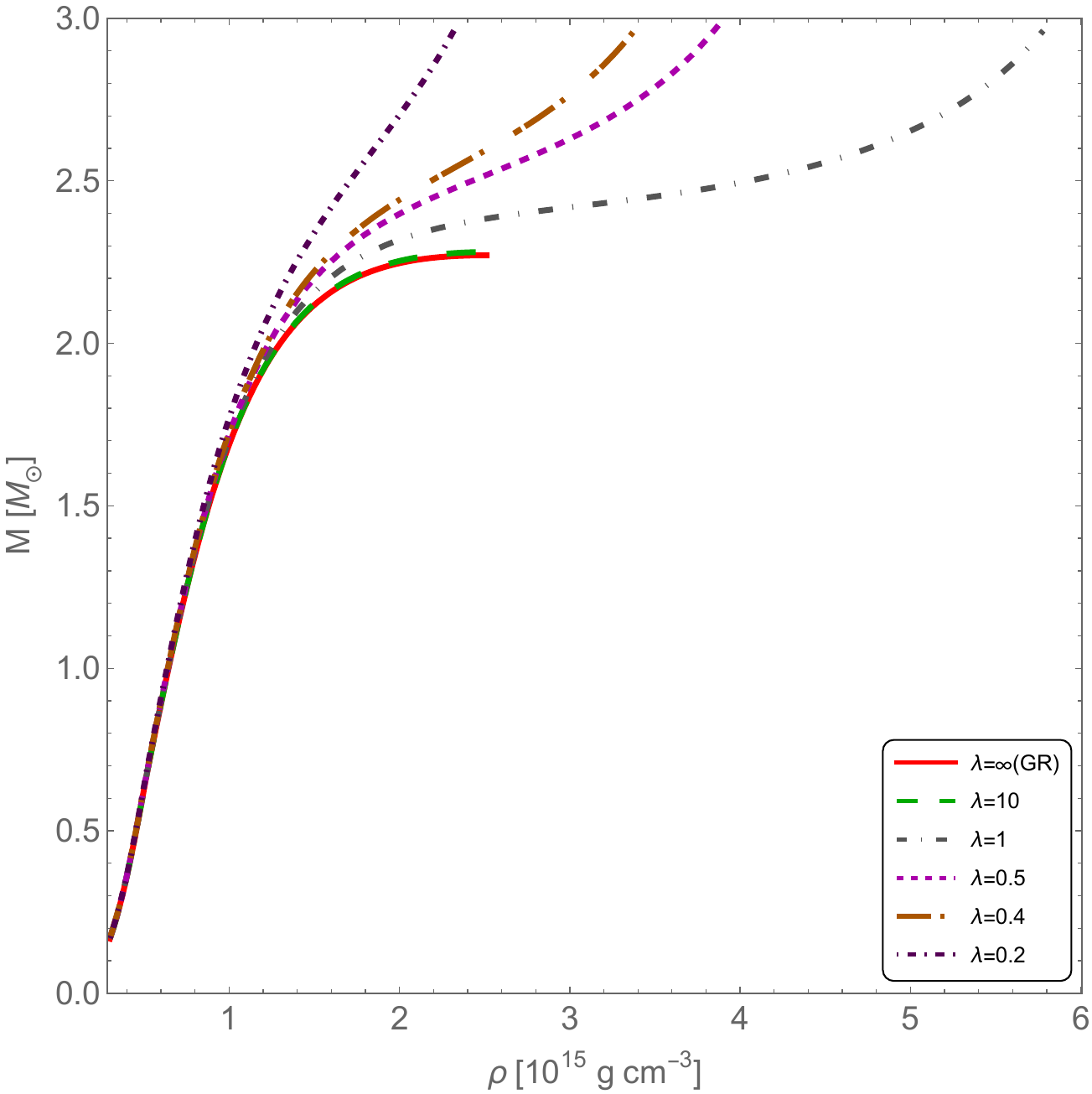}
    \caption{ENG}
    \label{fig:3c}
\end{subfigure}
\hfill
\begin{subfigure}{0.3\textwidth}
    \includegraphics[width=\textwidth]{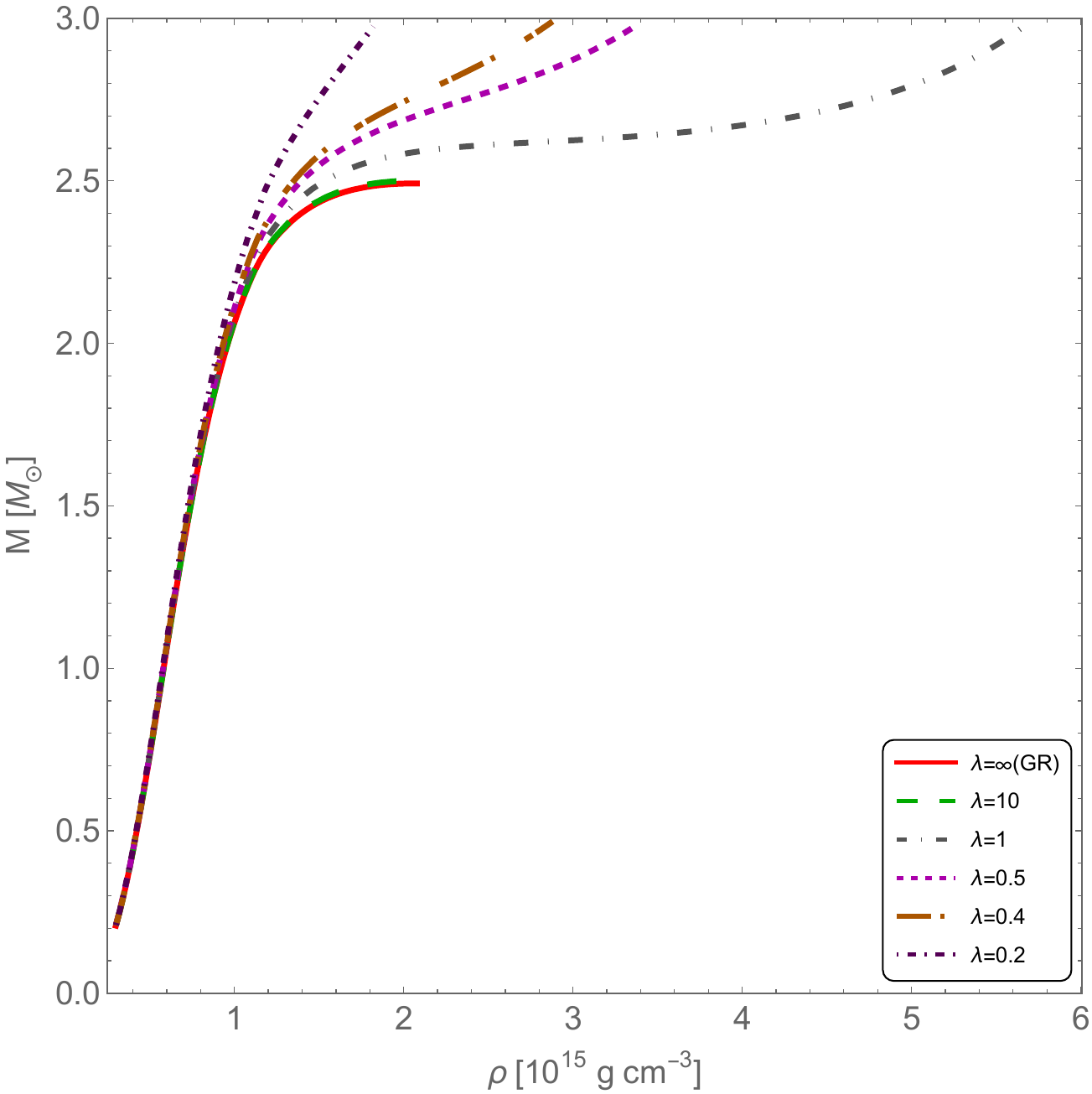}
    \caption{MPA1}
    \label{fig:3d}
\end{subfigure}
\begin{subfigure}{0.3\textwidth}
    \includegraphics[width=\textwidth]{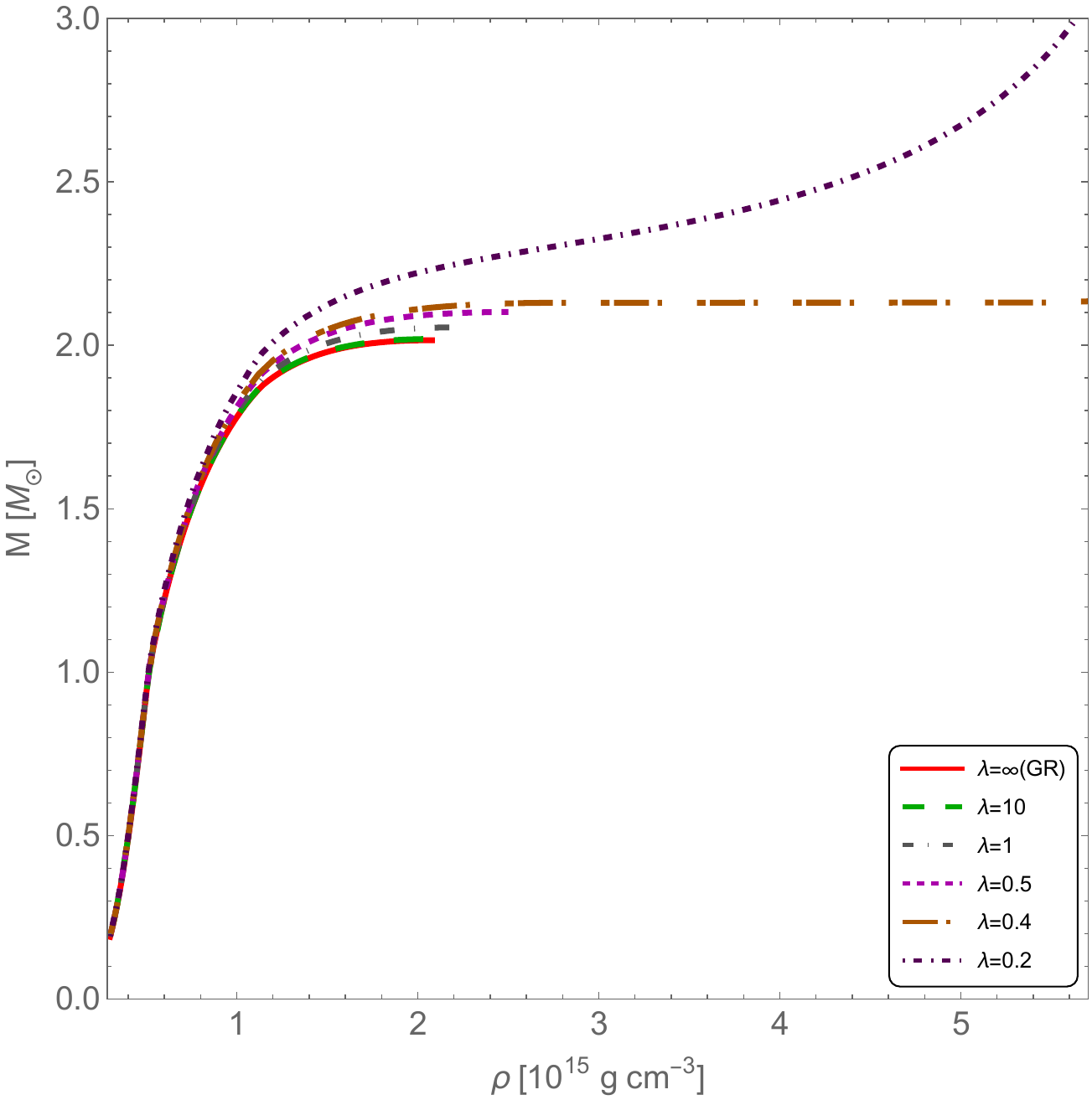}
    \caption{ALF2}
    \label{fig:3e}
\end{subfigure}
\begin{subfigure}{0.3\textwidth}
    \includegraphics[width=\textwidth]{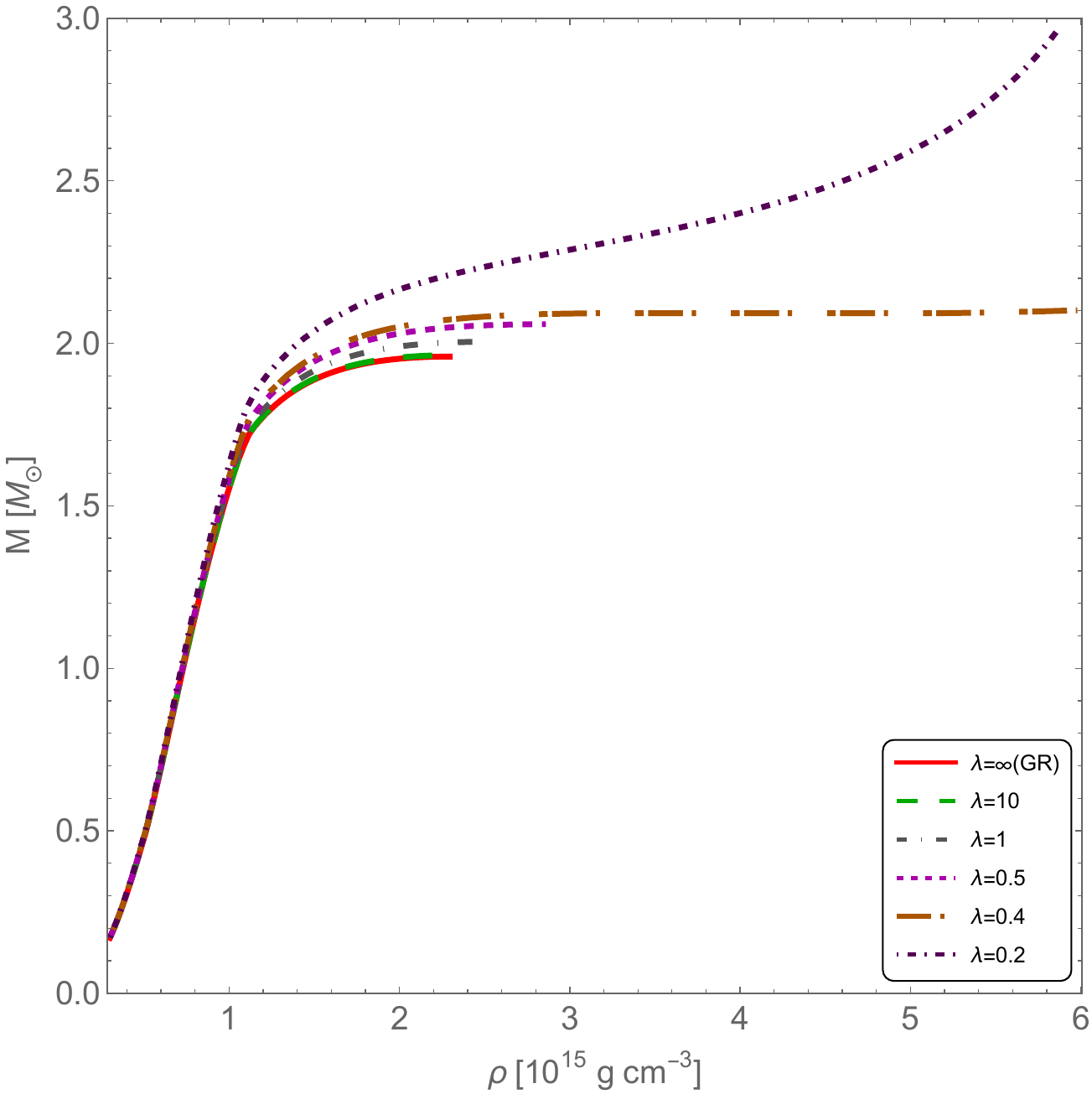}
    \caption{ALF4}
    \label{fig:3f}
\end{subfigure}
\caption{Mass-Density relation in braneworld scenario ($\lambda$ is given in units of  $10^{38}$ $\text{dyne}/\text{cm}^2)$.}
\label{fig:3}
\end{figure*}

\begin{figure*}[!ht]
\centering
\begin{subfigure}{0.3\textwidth}
    \includegraphics[width=\textwidth]{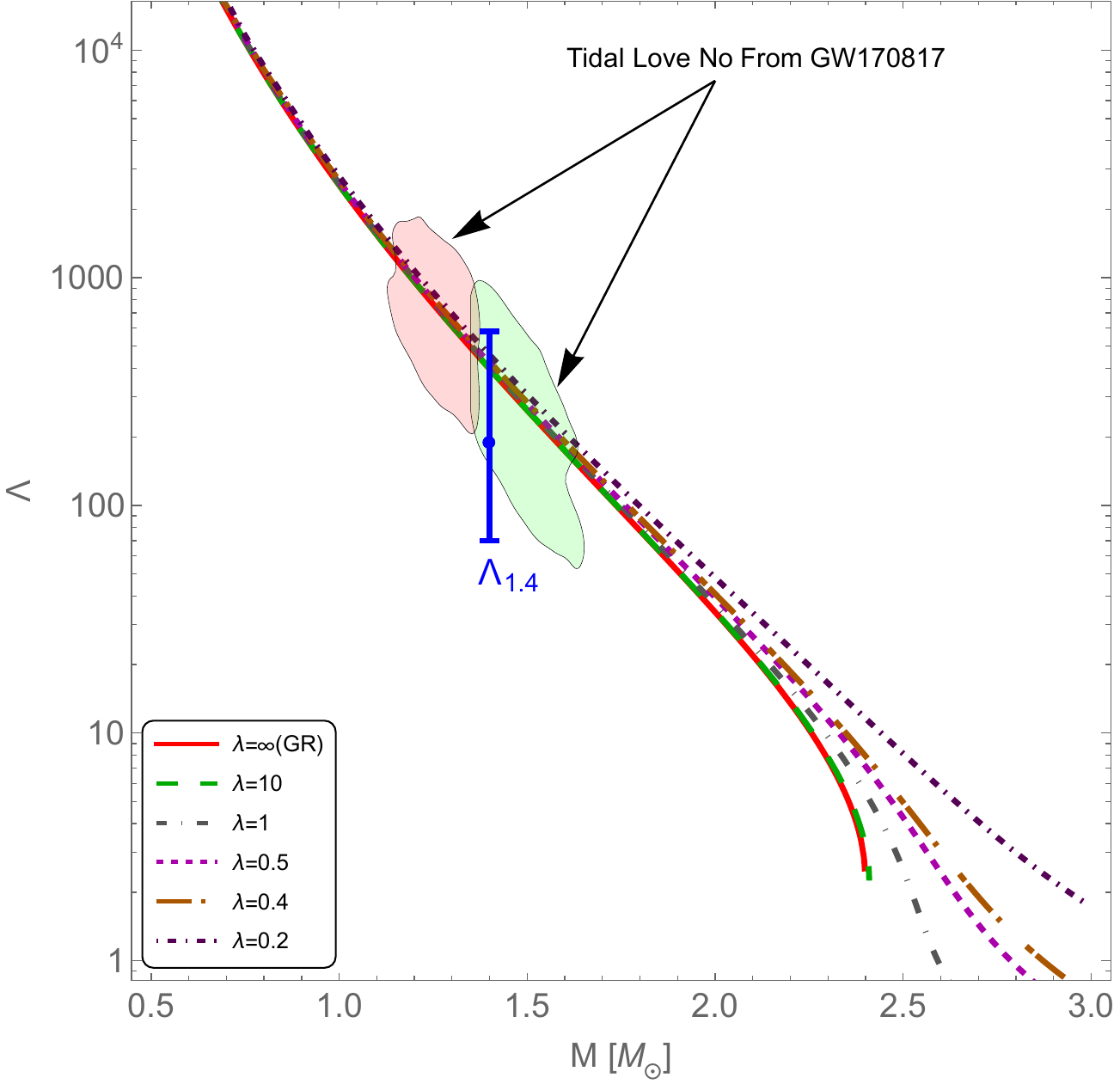}
    \caption{AP3}
    \label{fig:4a}
\end{subfigure}
\begin{subfigure}{0.3\textwidth}
    \includegraphics[width=\textwidth]{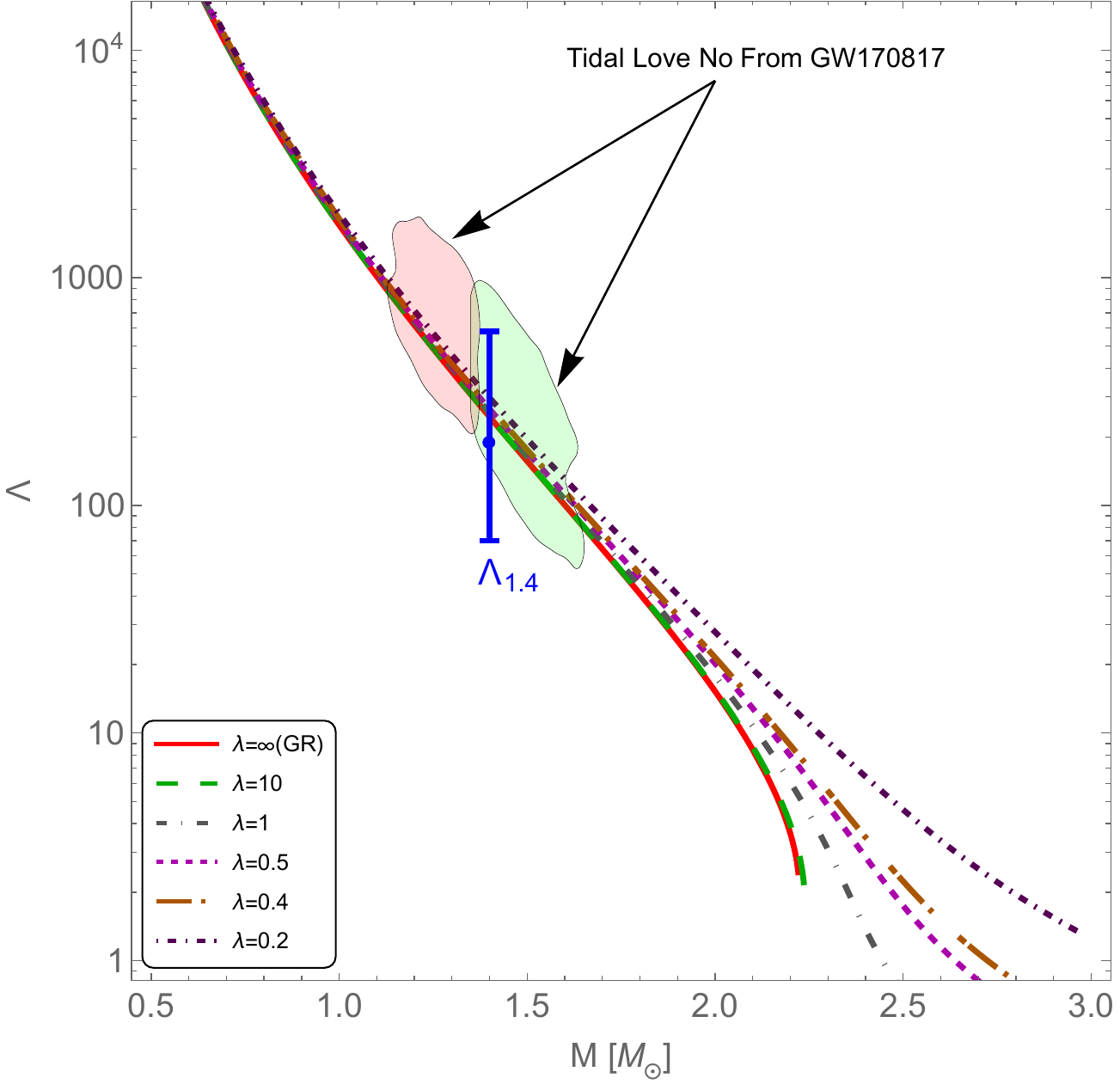}
    \caption{WFF2}
    \label{fig:4b}
\end{subfigure}
\begin{subfigure}{0.3\textwidth}
    \includegraphics[width=\textwidth]{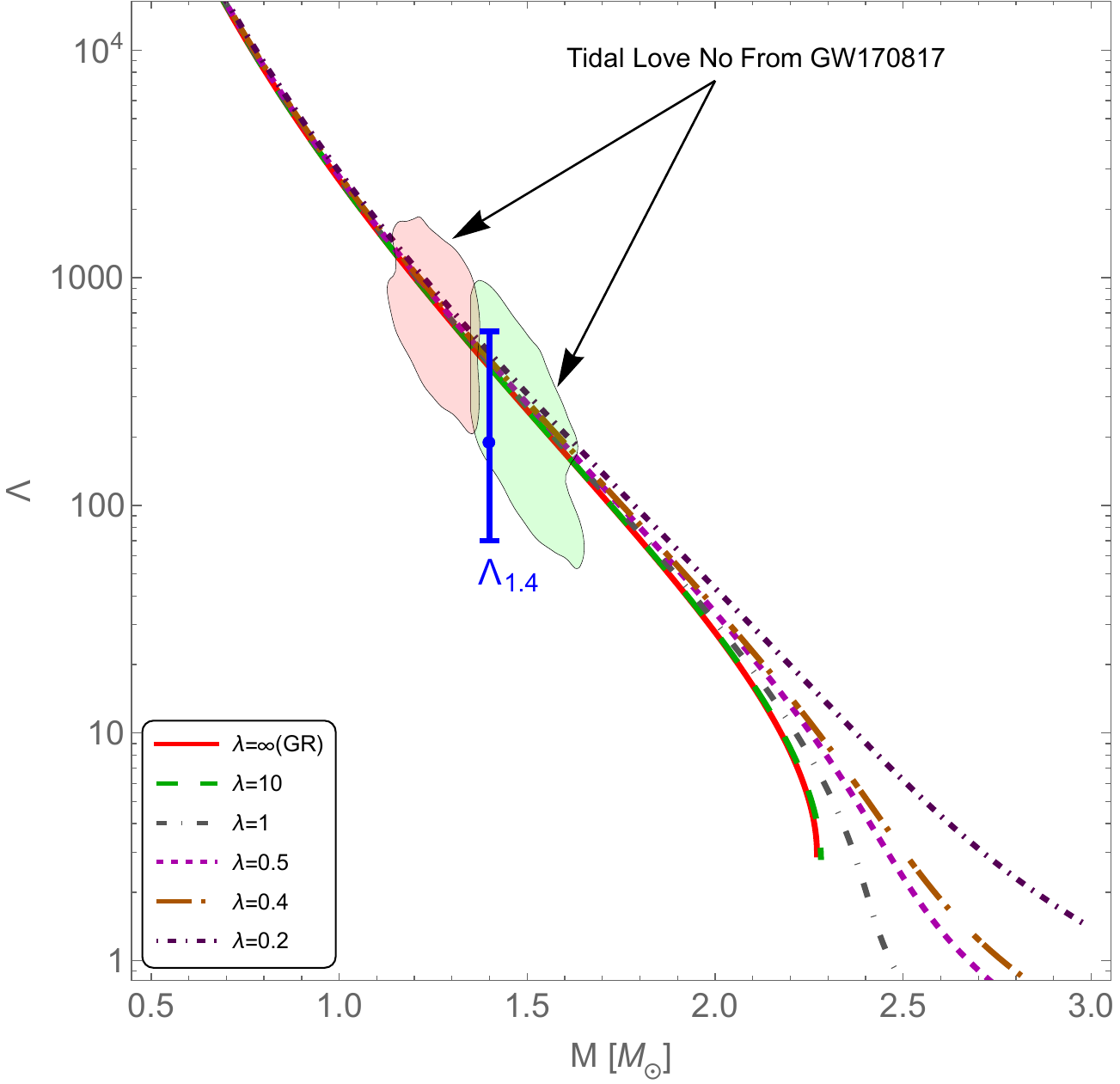}
    \caption{ENG}
    \label{fig:4c}
\end{subfigure}
\hfill
\begin{subfigure}{0.3\textwidth}
    \includegraphics[width=\textwidth]{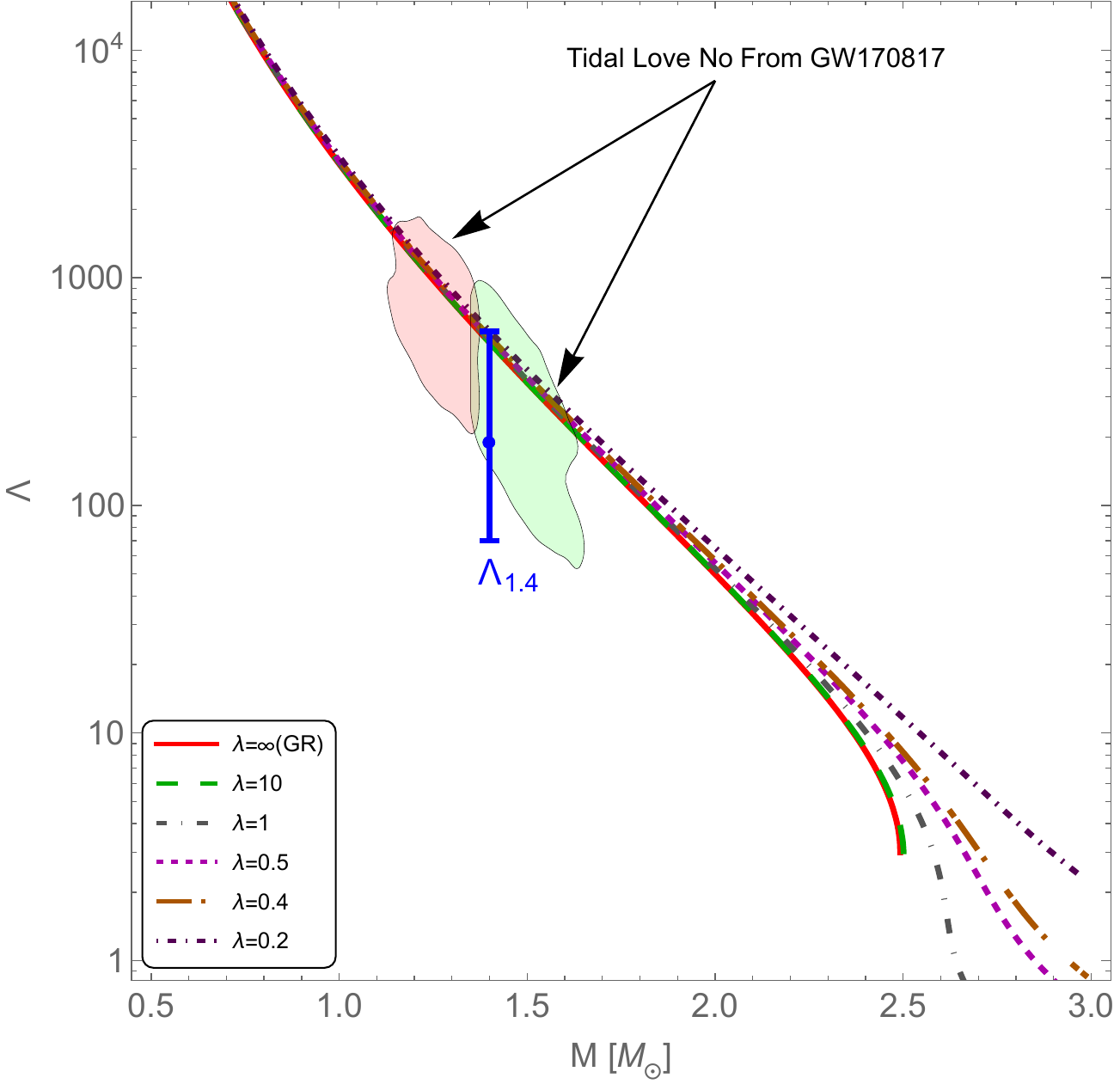}
    \caption{MPA1}
    \label{fig:4d}
\end{subfigure}
\begin{subfigure}{0.3\textwidth}
    \includegraphics[width=\textwidth]{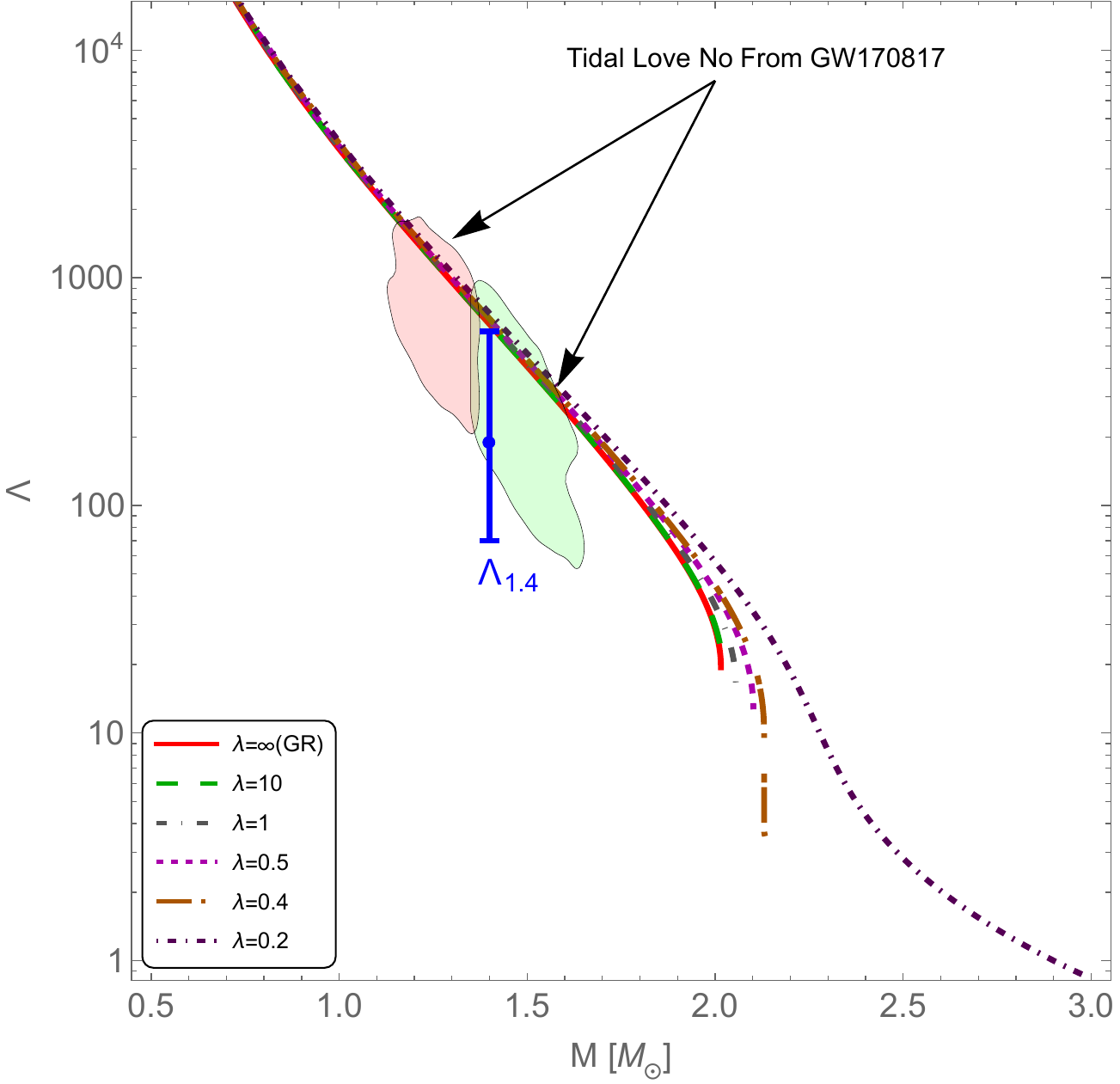}
    \caption{ALF2}
    \label{fig:4e}
\end{subfigure}
\begin{subfigure}{0.3\textwidth}
    \includegraphics[width=\textwidth]{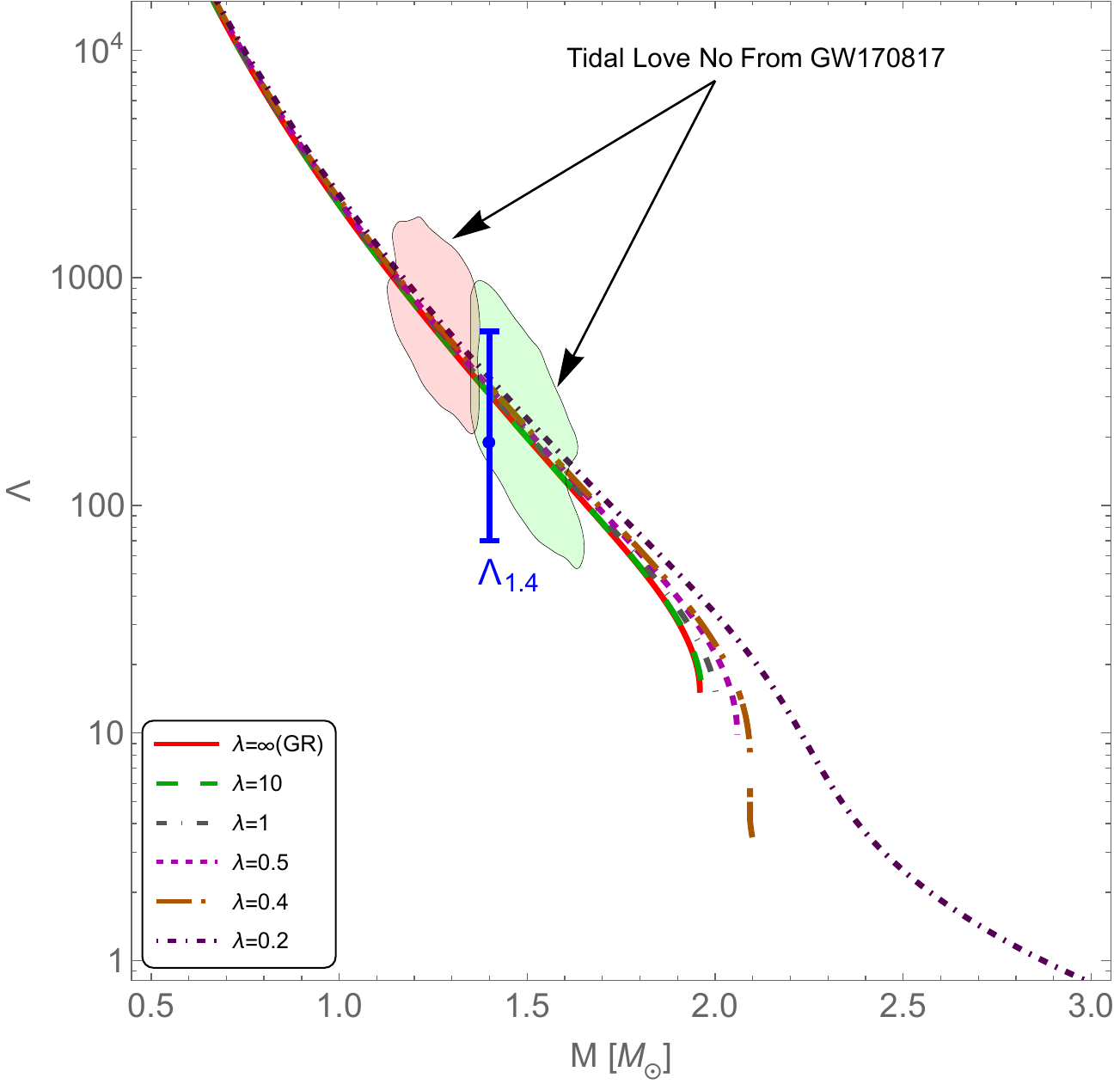}
    \caption{ALF4}
    \label{fig:4f}
\end{subfigure}
\caption{Mass-Love No. relation in braneworld scenario ($\lambda$ is given in units of $10^{38}$ $\text{dyne}/\text{cm}^2)$.}
\label{fig:4}
\end{figure*} 

%


\begin{table}[!ht]
    \centering
    \begin{tabular}{|c|c|c|c|c|c|}
      \hline
              EoS & \shortstack{$\lambda$ in \\ $10^{38}$ $\text{dyne}/\text{cm}^2$} &  R in km & \shortstack{Deviation in \\ radius from \\ GR  in \%} & $\Lambda$ & \shortstack{Deviation in \\ love no from \\ GR  in \%}\\ [1ex]
       \hline
        & GR  &  12.13 &  -  & 398  & - \\
        & 10  &  12.14 & 0   & 399 & 0.3 \\
        & 1   &  12.18 & 0.4 & 410 & 3.0 \\
   AP3  & 0.5 &  12.23 & 0.8 & 421 & 5.8 \\
        & 0.4 &  12.25 & 1.0 & 429 & 7.8 \\
        & 0.2 &  12.36 & 1.9 & 458 & 15.0\\
        \hline
        & GR  &  11.28 &  -  & 245 & - \\
        & 10  &  11.28 & 0   & 246 & 0.4  \\
        & 1   &  11.34 & 0.5 & 255 & 4.1 \\
   WFF2 & 0.5 &  11.40 & 1.0 & 265 & 8.2 \\
        & 0.4 &  11.42 & 1.3 & 270 & 10.0 \\
        & 0.2 &  11.57 & 2.6 & 297 & 21.0 \\
        \hline
        & GR  &  12.14 &  -  & 405 & - \\
        & 10  &  12.15 & 0   & 406 & 0.3 \\
        & 1   &  12.19 & 0.4 & 417 & 3.0 \\
   ENG  & 0.5 &  12.24 & 0.8 & 429 & 5.9 \\
        & 0.4 &  12.26 & 1.0 & 436 & 7.7 \\
        & 0.2 &  12.38 & 2.0 & 468 & 16.0 \\
        \hline
        & GR  &  12.60 &  -  & 518 & - \\
        & 10  &  12.60 & 0   & 520 & 0.4  \\
        & 1   &  12.64 & 0.3 & 531 & 2.5 \\
   MPA1 & 0.5 &  12.68 & 0.6 & 544 & 5.0\\
        & 0.4 &  12.70 & 0.8 & 550 & 6.2 \\
        & 0.2 &  12.80 & 1.6 & 584 & 13.0 \\   
        \hline
        & GR  &  11.68 &  -  & 307 & - \\
        & 10  &  11.69 & 0   & 309 & 0.7 \\
        & 1   &  11.74 & 0.4 & 319 & 3.9 \\
   ALF4 & 0.5 &  11.79 & 0.9 & 329 & 7.2 \\
        & 0.4 &  11.82 & 1.2 & 334 & 8.8 \\
        & 0.2 &  11.95 & 2.3 & 363 & 18.0 \\
        \hline
    \end{tabular}
    \caption{Radius and love No. (Round off to nearest Integer) of Canonical NS ($1.4M_{\odot}$) for different EoS in braneworld scenario with varying brane tension.}
    \label{tab:1}
\end{table}

\begin{table}[!ht]
    \begin{tabular}{|c|c|c|c|}
      \hline
              EoS & \shortstack{$\lambda$ in \\ $10^{38}$ $\text{dyne}/\text{cm}^2$} &  R in km & $\Lambda$\\ [1ex]
       \hline
        & 10   - & - & -  \\
        & 1   &  10.93 $\pm$ 0.3 & 1.57 $\pm$ 0.94 \\
   AP3  & 0.5 &  11.57 $\pm$ 0.1 & 2.97 $\pm$ 1.29 \\
        & 0.4 &  11.77 $\pm$ 0.1 & 3.57 $\pm$ 1.39 \\
        & 0.2 &  12.41 $\pm$ 0.0 & 6.31 $\pm$ 1.80 \\
        \hline
        & 10   - & - & - \\
        & 1   &  10.12 $\pm$ 0.06 & 0.56 $\pm$ 0.17 \\
   WFF2 & 0.5 &  10.74 $\pm$ 0.02 & 1.32 $\pm$ 0.42\\
        & 0.4 &  10.95 $\pm$ 0.02 & 1.71 $\pm$ 0.52 \\
        & 0.2 &  11.68 $\pm$ 0.05 & 3.68 $\pm$ 0.93 \\
        \hline
        & 10  - & - & -  \\
        & 1 &  10.28 $\pm$ 0.03 & 0.60 $\pm$ 0.20 \\
   ENG  & 0.5 &  11.05 $\pm$ 0.07 & 1.67 $\pm$ 0.65 \\
        & 0.4 &  11.31 $\pm$ 0.06 & 2.21 $\pm$ 0.82 \\
        & 0.2 &  12.11 $\pm$ 0.002 & 4.81 $\pm$ 1.39 \\
        \hline
        & 10  - & - & - \\
        & 1 &  11.50 $\pm$ 0.6 & 3.30 $\pm$ 2.54 \\
   MPA1 & 0.5 &  12.21 $\pm$ 0.2 & 5.18 $\pm$ 2.32\\
        & 0.4 &  12.38 $\pm$ 0.1 & 5.91 $\pm$ 2.33 \\
        & 0.2 &  12.91 $\pm$ 0.02 & 9.04 $\pm$ 2.61 \\   
        \hline
        & 10   - & - & - \\
        & 1    - & - & - \\
   ALF4 & 0.5  - & - & - \\
        & 0.4  - & - & - \\
        & 0.2 &  11.21 $\pm$ 0.07 & 2.02 $\pm$ 0.47 \\
        \hline
    \end{tabular}
    \caption{Radius and love No. of GW190814 NS having mass around  $2.5-2.67$ $M_{\odot}$ for different EoS in braneworld scenario with varying brane tension.}
    \label{tab:2}
\end{table}

\section{Results and discussions}

To solve equations \ref{S2EQ:7}-\ref{S2EQ:10}, we require an equation of state (EoS). Numerous EoSs are documented in the literature. In this paper, we employ a piece-wise polytropic (PWP) approach. In J.S. Read's pioneering work, there are 34 PWP EoSs available, of which only AP3, WFF2, ENG, MPA1, ALF2, and ALF4 satisfy the constraints outlined in GW170817, specifically the mass-radius and mass-tidal deformability (see fig. \ref{fig:1}) constraints, as well as those for PSR J0740 and PSR J0030 in the context of GR.

Applying the canonical neutron star (NS) radius of $ R_{1.4} = 10.9^{+1.9}_{-1.5} $ at a 90\% confidence level \cite{Bharat:2019} and the canonical NS tidal deformability of $\Lambda_{1.4} = 190^{+390}_{-120} $ at a 90\% confidence level \cite{Abbott:2018}, it becomes evident that the ALF2 EoS fails to satisfy these two criteria regarding the canonical NS radius and tidal deformability within the framework of GR. Furthermore, these EoSs do not align with the predicted mass for NSs exceeding $ 2.5 M_{\odot} $ in GR, as indicated by GW190814.

Lugones and Arbanil employed the casual limit EoS within the Braneworld model, showing that the maximum mass limit set by GR can be exceeded. In line with their approach, we apply the six PWP mentioned earlier EoSs within the Braneworld framework to investigate whether we can constrain the Brane tension parameter using observational data and get NS with mass that of mass predicted by  GW190814.

In Fig.-\ref{fig:2}, we present the mass-radius plot for the previously mentioned EoSs. In the braneworld scenario, the brane tension exerts a more pronounced effect on NSs that exceed the canonical mass. For a given mass, the radius increases as the brane tension decreases. This change in radius due to the increase in $\lambda$ is more evident in heavier masses than in lighter masses. For each EoS, the mass-radius plot exhibits a turning point, a significant feature that piques curiosity. Beyond this turning point, the mass-radius plot transitions to a linear pattern. For the AP3, WFF2, ENG, and MPA1 EoS, this turning point occurs at a brane tension of $ \lambda = 10^{38} \, \text{dyne} / \text{cm}^{2}$. In contrast, for the ALF2 and ALF4 EoS, the turning point is observed at $ \lambda = 2 \times 10^{37} \, \ \text{dyne} / \text{cm}^{2} $. This suggests that the turning point occurs at a higher value of brane tension for stiffer EoS compared to softer EoS. We consider central density up to those values for which $\dfrac{dM}{d\rho}>0$ is satisfied. The maximum central density increases with the decrease in brane tension (see Fig.-\ref{fig:3} ). In Braneworld, as brane tension decreases, the Weyl energy density causes neutron stars to have increasingly higher central densities, resulting in NSs that are heavier than those predicted by GR. In Fig.-\ref{fig:4}, we depict the relationship between mass and tidal deformability within the braneworld scenario. The deviation in mass-tidal deformability relation from its GR counterparts is less noticeable for neutron stars with masses below the canonical value. As the brane tension decreases, the NSs exhibit more significant tidal deformation.

Each EoS was consistent with the observational data from GW170817, PSR J0740, and PSR J0030 within the context of the braneworld scenario. The canonical NS radius, $ R_{1.4} = 10.9^{+1.9}_{-1.5} $ and tidal deformability of $\Lambda_{1.4} = 190^{+390}_{-120} $  at a 90\% confidence level, emerged as a pivotal tool in constraining the braneworld tension. Its robustness in providing a more stringent restriction than the combined constraints derived from GW170817, PSR J0740, and PSR J0030 cannot be overstated.

 It is crucial to highlight that the ALF2 EoS, even in the braneworld scenario, fails to meet the canonical NS radius and tidal deformability constraints. The mass-radius relationship of the MPA1 EoS, given \( \lambda = 2 \times 10^{37} \, \text{dyne} / \text{cm}^{2} \), approaches the upper limit of the canonical neutron star radius. Similarly, the mass-tidal deformability relation for the MPA1 EoS, under the same parameter \( \lambda \), exceeds the maximum value of canonical tidal deformability. These two relations for the MPA1 EoS set the upper limit of the brane tension. In Table-\ref{tab:1}, we illustrate the extent to which the radius and tidal deformability differ from their GR counterparts. It indicates that tidal deformability exhibits a greater deviation from its GR counterparts compared to the radius for a specified brane tension. In Table-\ref{tab:2}, we present the predicted radius and tidal deformability range of the GW190814 NS for various values of the brane tension. It illustrates that for the EoSs AP3, WFF2, ENG, and MPA1, we can attain GW190814 NS mass with a brane tension of \( \lambda = 1 \times 10^{38} \, \text{dyne} / \text{cm}^{2} \). In contrast, for the ALF4 EoS, reaching the same GW190814 NS mass requires a brane tension of \( \lambda = 2 \times 10^{37} \, \text{dyne} / \text{cm}^{2} \).

\section{Conclusions}
In this study, we investigated the properties of NSs within the Braneworld model by utilizing six distinct PWP EoSs. These EoSs adhere to the mass-radius and mass-tidal deformability constraints established by GW170817 and observations from pulsars PSR J0740 and PSR J0030. Subsequently, we utilized the canonical NS radius and tidal deformability to constrain the brane tension parameter $\lambda$. Our primary objective was to assess whether these EoSs, in conjunction with the Braneworld framework, could support the existence of more massive neutron stars, as suggested by the GW190814 event, while remaining consistent with the aforementioned observational constraints.

The brane tension affects NSs more with masses greater than the canonical mass. As the brane tension lessens, the radius and the tidal deformability of these NSs increase, with these effects being more significant for the heavier stars. The mass-radius curve features a distinct turning point for each EoS, after which the plot becomes linear. The location of this turning point depends on the stiffness of the EoS, with stiffer EoSs (such as AP3, WFF2, ENG, and MPA1) showing a higher brane tension at the turning point compared to softer EoSs (ALF2 and ALF4).

The canonical NS radius and tidal deformability constrain the brane tension parameter, $\lambda$. Our analysis reveals that under the ALF2 EoS, the canonical constraints on the NS radius and tidal deformability are not satisfied, even within the Braneworld scenario. This suggests that certain softer EoSs, like ALF2, are incompatible with the observed relationships between mass-radius and mass-tidal deformability. By comparing the canonical NS radius and tidal deformability with the results obtained for each EoS within the Braneworld framework, we can establish strong constraints on the brane tension parameter. Notably, the MPA1 EoS, which presents a mass-radius relationship that nearly reaches the upper limit of the canonical NS radius and tidal deformability, provides a stringent lower bound on the value of the brane tension to be $\lambda < 2 \times 10^ {37} \, \text{dyne} / \text{cm}^{2} $. The constraint on the brane parameter presented herein demonstrates a considerably greater strength than those derived from the Big Bang cosmology and the constraints obtained through some astrophysical investigations \cite{Castro:2014, Felipe:2016}. Nevertheless, it is essential to note that this remains a comparatively less stringent limitation imposed by experimental studies focused on Newtonian gravitational equations \cite{Maartens:2000b}.

Our results confirm that, within the Braneworld scenario, the maximum NS mass can exceed the values predicted by GR, aligning with the mass of the NS observed in the GW190814 event. The necessary brane tension values to attain this elevated mass differ among the EoSs, contingent upon the stiffness of each EoS. For instance, the EoSs AP3, WFF2, ENG, and MPA1 necessitate a brane tension of $ \lambda = 10^{38} \, \text{dyne} / \text{cm}^{2}$ to reach the NS mass associated with the GW190814 event, whereas the softer ALF4 EoS requires a lower brane tension of $ \lambda = 2 \times 10^{37} \, \text{dyne} / \text{cm}^{2}$.

In conclusion, this study highlights the Braneworld model, which aligns with recent observations and can be used to explain neutron stars with masses surpassing those suggested by GR. It also highlights the significance of the brane tension parameter on NS characteristics.

\section*{Acknowledgements}
MK is grateful to ICARD, Aliah University for providing research facilities and Inter-University Centre for Astronomy and Astrophysics (IUCAA), Pune, India for providing Associateship programme under which a part of this work was carried out. The authors express their gratitude to the referee for providing valuable suggestions.

\bibliographystyle{apsrev}
\bibliography{reference}

\end{document}